\documentclass[preprint, 11pt]{elsarticle}
\graphicspath{{figures/}}

\usepackage{lineno,hyperref}
\usepackage{subcaption}
\usepackage[dvipsnames]{xcolor}
\usepackage{placeins}
\usepackage{multirow}
\usepackage[normalem]{ulem}
\modulolinenumbers[5]

\journal{Chaos, Solitons and Fractals}

\begin{document}

\begin{frontmatter}

\title{Temporal Taylor's scaling of facial electromyography and electrodermal activity in the course of emotional stimulation}

\author[PW]{Jan Cho\l oniewski\corref{jch}}
\ead{choloniewski@if.pw.edu.pl}
\author[PWr]{Anna Chmiel}
\author[PW,MP]{Julian Sienkiewicz}
\author[PW,ITMO]{Janusz Ho\l yst}
\author[JU]{Dennis K\" uster}
\author[JU]{Arvid Kappas}

\address[PW]{Faculty of Physics, Center of Excellence for Complex Systems Research, Warsaw University of Technology, Koszykowa 75, PL00662 Warsaw, Poland}
\address[PWr]{Department of Theoretical Physics, Wroc\l aw University of Technology, Wybrze\. ze Wyspia\' nskiego 27, PL50370 Wroc\l aw, Poland}
\address[MP]{Max Planck Institute for the Physics of Complex Systems, N\" othnitzer Strasse 38, DE01187 Dresden, Germany}
\address[JU]{Department of Psychology and Methods, Jacobs University, Campus Ring 1, DE28759 Bremen, Germany}
\address[ITMO]{ITMO University, Kronverkskiy av. 19, RU197101 Saint Petersburg, Russia}

\cortext[jch]{corresponding author}

\begin{abstract}
High frequency psychophysiological data create a challenge for quantitative modeling based on Big Data tools since they reflect the complexity of processes taking place in human body and its responses to external events. Here we present studies of fluctuations in facial electromyography (fEMG) and electrodermal activity (EDA) massive time series and changes of such signals in the course of emotional stimulation. \textit{Zygomaticus major} (ZYG; ``smiling'' muscle) activity, \textit{corrugator supercilii} (COR; ``frowning'' muscle) activity, and phasic skin conductance (PHSC; sweating) levels of $65$ participants were recorded during experiments that involved exposure to emotional stimuli (i.e., IAPS images, reading and writing messages on an artificial online discussion board). Temporal Taylor's  fluctuations scaling were found when signals for various participants and during various types of emotional events were compared. Values of scaling exponents were close to 1, suggesting an external origin of system dynamics and/or strong interactions between  system's basic elements (e.g., muscle fibres). Our statistical analysis shows that the scaling exponents enable identification of high valence and arousal levels in ZYG and COR signals.
\end{abstract}


\end{frontmatter}

\section{Introduction} 
Easy access to massive amounts of high frequency data about humans --- their health~\cite{ania_njp,klimek} and their responses (also remote)~\cite{ania_physa, ania_plos,entropy,choloniewski} --- is an important fruit of the so-called {\it Big data} science~\cite{bigdata1,bigdata2,bigdata3}. In this scope psychophysiological information that can be transformed into undisputed facts/relations concerning our vitals organs~\cite{urbanowicz} is of utmost importance. 

In 1961, ecologist Lionel Roy Taylor published his famous paper~\cite{taylor} in which he reported a power-law relation between a sample variance of density and a mean density of a sample of several species in a study area. The data was taken from observations of many  species, e.g., various kinds of larvae, worms, symphylas, macro-zooplankton, shellfish, etc. Taylor was considering the scaling exponents as a universal measure for aggregation of population abundance. In his opinion~\cite{taylor}, the strong aggregation should correspond to a larger scaling exponent, and it should be a result of mutual attractions between individuals belonging to a given species. Mutual repulsion should result in lowering of spatial dispersion and in lowering of the scaling exponent.

In fact, a similar scaling law was found already in 1938 by statistician H. Fairfield Smith who described it in the (often forgotten) paper \cite{smith} entitled {\it  An empirical law describing heterogeneity in the yields of agricultural crops}.  Smith compared yields of wheat, maize, sorghum, mangolds and  potatoes from different areas and found that {\it the regression of the logarithms of the variances for plots of different areas on the logarithms of their areas was approximately linear} \cite{smith}. Slopes of the regression curve (we call them scaling exponents)  varied from crop to crop and even from one plant's region to another. However they were always smaller than a value received for  uncorrelated plants and Smith connected a specific value of  regression slope   with the crop heterogeneity.

It is likely that the first observation of the power-law relation  between between a mean  and a variance  was found by C.I. Bliss \cite{bliss1941}  who studied  variations in populations density of  Japanese beetle larvae in 1941\footnote{We are thankful to an anonymous Referee for suggesting us this reference.}.    

In general, Taylor's theorem leads to relating the standard deviation   of an additive variable with its mean value in similar systems as: $ \sigma_i\propto \langle f_i\rangle^\alpha$
where $\sigma_i$ -- a standard deviation of a given additive value for $i$-th subsystem, $\langle f_i\rangle$ -- a mean of the value. 
If the scaling exists then the value of the exponent $\alpha$ allows to infer about underlying dynamics of the analyzed system by comparing the results with a behavior of agent-based models or stochastic  dynamics \cite{kertesz,tadic2007transport,menezes2004B}. 

Studies of Smith and Taylor were devoted to observations of variance between samples of some yields, or number of some animals occupying different areas of the same surface. The proposed relationship was later confirmed in several other empirical studies in ecology (see for example Ref.~\cite{elliott2003,mcardle1990}), life sciences (e.g., scaling of cell numbers in representatives of a given species~\cite{azevedo}), astrophysics~\cite{uttley}, company  growth rates~\cite{amaral} or the stock market \cite{kertesz}. For more examples and for theoretical models that try to explain the power law between the variance and the mean see, e.g.,  review papers  \cite{kertesz,meng}.

The above mentioned empirical studies considered so-called \textit{ensemble fluctuations scaling}  (EFS) since variances were calculated over  an ensemble of samples (subsystems)  belonging to the same class (usually the class was labeled by a given surface of samples). EFS can be  called also  a  {\it spatial Taylor's law} since in ecology it shows  how populations vary in spatial aggregation as a function of their average size. It means  points at the scaling plot are described  by   means and  variances of a set of spatially distinct locations within a population.

There is also another kind of scaling called  \textit{temporal fluctuations scaling}  that, in ecological systems, relates the variability of populations  {\bf time-series} to their mean \cite{keitt2002}. In such a case each point at the scaling plot corresponds to  a mean and a variance of a single population time series. This kind of scaling was also observed in many  natural and man-made systems \cite{kertesz}, including various kind of networks such as internet routers, river networks, highways networks or World Wide Web  \cite{menezes2004A}. 
 
As far as we know  at the moment there is no common agreement on reasons of observed Taylor's scaling and there are several theories trying to explain this effect.  Temporal scaling  in ecology can be for example results of  environmental and demographic stochasticity  \cite{Ballantyne2005b, Ballantyne2007} or  interspecific competition \cite{kilpatrick2003}. There are also attempts to find domain-independent {\bf general} roots of the Taylor's scaling, e.g., in probabilistic models known as the Tweedie exponential dispersion models  that follow from  central-limit-theorem-like theorem  \cite{kendal2011} or i.i.d. processes with skewed distributions  \cite{cohen2015}.  

A natural question is whether the origin of the observed temporal fluctuations and the scaling law is an effect of a stochastic external driving force or the randomness of complex system internal dynamics \cite{kertesz,keitt2002}.  In \cite{menezes2004B}, it was suggested that one can separate both contributions and estimate the ratio of internal interactions between the system’s components and the influence of external perturbations. Studies of temporal scaling  for fluctuations of traded values at NYSE and NASDAQ stock markets \cite{kertesz,eisler} have shown that the scaling exponent strongly depends on the length of the time window where the variability was observed, and this dependence can provide information about correlations taking place in the system in different time scales \cite{kertesz,keitt2002}. Comparative investigations  of fluctuations scaling of quotation activity at an online foreign exchange market (Forex) are presented in \cite{Aki1} and \cite{Aki2}. Theoretical studies aiming to build agent-based models explaining the temporal scaling can be found, e.g., in \cite{tadic2007transport} or \cite{duch}, where various kinds of network topology and random walks were considered.  

The main focus of this paper is to study high frequency fluctuations in facial electromyography (fEMG) and electrodermal activity (EDA) time series in the course of various emotional stimulation episodes. Our goal is to check if biological subsystems sensitive to human emotions, i.e. facial muscles responsible for smiling (\textit{zygomaticus major}) and frowning (\textit{corrugator supercilii}) expressions \cite{larsen}, and skin sweat glands (their activity can be associated with a human's arousal \cite{bach2}) follow the temporal Taylor's scaling. Mean values and variances of the signals during visual emotional stimulation have been calculated in time windows of different sizes. To the best of our knowledge, the presence of such a scaling was never reported for psychophysiological signals. The challenge is to distinguish between various human emotions using observations of the temporal scaling of Taylor's law.  

\section{Description of temporal Taylor's fluctuation scaling}
In the present paper, we will consider temporal Taylor's scaling \cite{kertesz,tadic2007transport,menezes2004B,keitt2002,menezes2004A,duch}. 
Let $f_{i,t}$ be a positive variable $f_{i,t}$ describing an additive measure of a given activity of the object $i$ at time moment $t$.  Examples of such activities can be  data packages coming to a router, or visits of a web page or activated muscle fibers.  Let the total number of elements  in time series of this activity be $T$, i.e. $t=1,2,3, \dots, T$ (further  we will assume that $T$ is the same for all objects $i$). Let us divide the series into $Q$ windows of size $\Delta$, i.e., $Q \Delta =T$. 
The quantity $[f_i^{(q,\Delta)}]$ stands for a cumulative value of the variable $f_i$ in a  window of the size $\Delta$ ($q=1,2,3, \dots,Q$ is the window's label) and $(\sigma_i^{(\Delta)})^2$ is a variance of this cumulative variable in the whole data series. 
Then we have 
\begin{equation}
(\sigma_i^{(\Delta)})^2= \left\langle[f_i^{(q,\Delta)}]^2\right\rangle- \left\langle[f_i^{(q,\Delta)}]\right\rangle^2
\end{equation}
Here 
\begin{equation}
 \left\langle[f_i^{(q,\Delta)}]\right\rangle = \frac{1}{Q}\sum_{q=1}^{Q}\sum_{t=(q-1)\Delta+1}^{q\Delta}f_{i,t}= \Delta\frac{\sum_{t=1}^{T}f_{i,t}}{T}
\end{equation}
and
\begin{equation}
 \left\langle[f_i^{q,\Delta}]^2\right\rangle=\frac{1}{Q}\sum_{q=1}^{Q}\left(\sum_{t=(q-1)\Delta+1}^{q\Delta}f_{i,t}\right)^2
\end{equation} 
When the window $\Delta$ is kept constant for all objects $i$ belonging to a given system (e.g. a network  of Internet routers or WWW)  then Taylor's scaling means  
\begin{equation} 
 \sigma_i^{(\Delta)}\propto\left\langle[f_i^{q,\Delta}]\right\rangle^{\alpha(\Delta)}. \label{Taylor}
\end{equation}

The  value of the exponent $\alpha(\Delta)$ can be dependent on the window size $\Delta$ \cite{kertesz,duch}, and such a dependence can bring some additional information on dynamics of constituents forming a considered system. The case $\alpha(\Delta)=1/2$ can correspond to a system consisting of {\it mutually independent elements} as in the case of the ideal gas or for random  processes obeying the  Central Limit Theorem\footnote{Let us note that for the  ensemble fluctuation scaling there are i.i.d. processes with skewed distributions  that can lead to other values of  the parameter $\alpha$, see \cite{cohen2015}}. This value can also be observed when the variable $f_i$ corresponds to a number of some events  (e.g., data packages coming to a given node $i$)  and  when  the time window $\Delta$ is so short that it is very unlikely that more than two events can emerge in a single window \cite{kertesz}. Larger values of the exponent  $\alpha(\Delta)$ can correspond to a larger  degree of synchronization of elements forming the system and  a set of completely synchronized elements displays scaling $\alpha(\Delta)=1$. Let us stress that synchronization  in ecological systems  is frequently observed, see e.g., coupling of trees reproduction cycles  via pollen exchange \cite{satake2000,Ballantyne2005}. A similar situation takes place when a system is driven by an external force, for example when populations of separated groups of animals are synchronized by weather conditions \cite{morgan1953} or by other environmental influences \cite{Grenfell1998}.

A possible explanation of exponents $\alpha(\Delta) > 1$ can be provided in several ways, e.g., by Tweedie distributions and impact inhomogeneity \cite{kertesz}. Bar-Lev and Enis have proved \cite{barlev1986} that such exponents can be also observed when the activity $f_{i,t}$ is governed by so-called stable distributions with characteristic exponents between 0 and 1.

Observations  of Taylor's scaling should not be confused with  \textit{Detrended Fluctuation Analysis} (DFA)  \cite{dfa}  and other approaches~\cite{eke2000,eke2002} used to quantify long-range power-law correlations in various signals~\cite{ckpeng} including psychophysiological data related to emotional states \cite{DFA6,DFA8,DFA9,dfa_emocje1,dfa_emocje2}. Let us stress then while the DFA framework uses a single time series the Taylor's scaling approach describes properties of a set of similar objects (even when they are independent). It follows that the exponent $\alpha'$  of DFA scaling and the exponent  $\alpha(\Delta)$ of Taylor's scaling defined in Eq. (\ref{Taylor}) are not the same although there exists a relation between their derivatives
\[
\frac{\mathrm{d}\alpha'}{\mathrm{d}(\log{\left\langle f_i\right\rangle})}\sim \frac{\mathrm{d}\alpha(\Delta)}{\mathrm{d}\log{\Delta}} 
\]
where $\left\langle f_i\right\rangle$ is the mean value of the analyzed signal. For a derivation of the relation and more information - see Ref. \cite{kertesz}.

\section{Data}\label{sec:data}
\begin{figure}[tb]
\centering
\includegraphics[width=\textwidth]{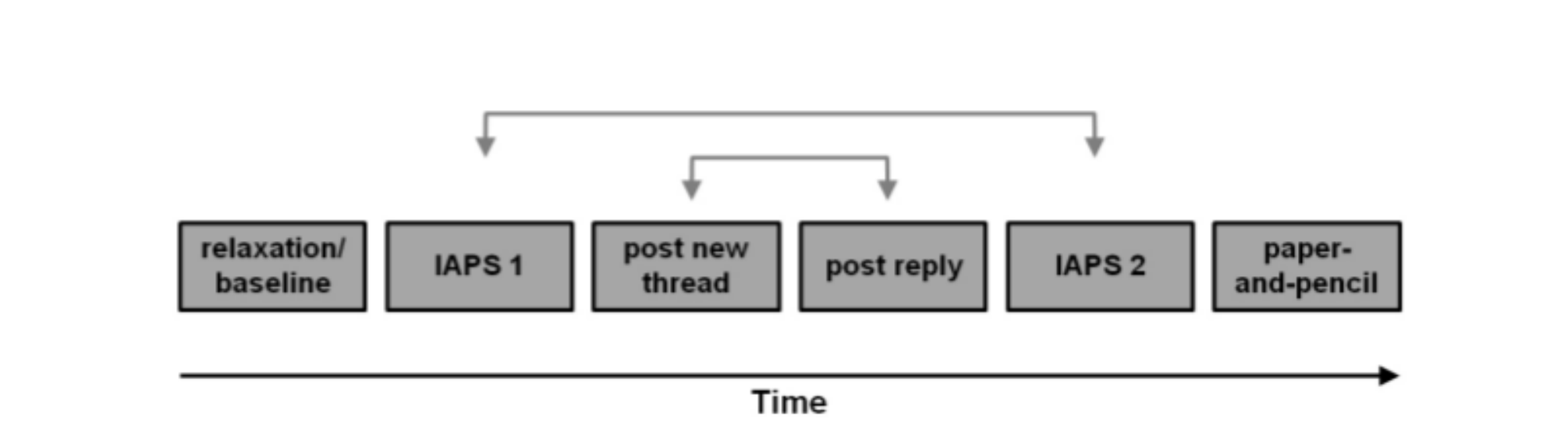}
\caption{\textbf{Graphic representation} of the sequence of the main experimental blocs of the study. The arrows indicate randomizations of blocks between participants. ``Relaxation/baseline'' -- participants were asked to relax for 3 minutes; ``IAPS 1'' and ``IAPS 2'' -- two rounds of IAPS images presentation, 4 images in each round, sequence for each image: $2s$ of blank image, $6s$ of IAPS exposition, emotion questionnaire; ``post new thread''/``post reply'' -- parts of experiment not related to the study; ``paper-and-pencil'' -- final questionnaire regarding demographics (not used in the study).}
\label{fig:eksperyment}
\end{figure}

Our data were gathered during an experiment conducted at Jacobs University Bremen, Germany. There were 65 participants (30 female; mean age = 20.4 years; standard deviation in the sample = 1.9) that were subjected to emotional stimuli -- pictures from the International Affective Picture System (IAPS) \cite{lang} and forums (see Fig.~\ref{fig:eksperyment} for schematic representation of the experiment). During the course of the experiment, participants' fEMG (\textit{corrugator supercilii}, \textit{zygomaticus major}) was recorded with the sampling frequency $\nu_s = 2\mathrm{kHz}$, using the BIOPAC MP150 amplifier system (Biopac Systems, Inc., Santa Barbara, CA), and signal amplitudes ($\mu$V) were amplified with a gain factor of 5000. EDA (left and right foot skin conductance) was recorded with the same system at a rate of 500Hz, and an amplification of 5$\mu$S/V. \textit{Corrugator supercilii} refers to the main facial muscles controlling eyebrow movements, such as frowning; \textit{zygomaticus major} refers to the facial muscles primarily responsible for raising lip corners in smiling. Markers were placed in the time series to allow identification of an event taking place at a given time. The total volume of the considered dataset was around  $25$ GB. 

Emotions elicited during IAPS image presentations were scored by the participants in questionnaires with three questions asking separately about experienced \textit{positive}, \textit{negative} emotion and \textit{arousal} on Likert-type scales from 1 to 7. The basis for this assessment was a dimensional theory of emotions that focuses on ``single simple feelings'' \cite{Yik2011}, i.e. ``Core Affect'' that can be represented, or mapped, onto only two or three core dimensions \cite{Zeng2009}, whereas other emotion theories such Ekman's \textit{Neurocultural Theory} \cite{ekman1994} distinguish a small number of categorically distinct emotions such as anger, fear, sadness, happiness, disgust, or surprise. One of the basic assumptions of dimensional models is that valence and arousal are primary and automatically perceived, whereas categories are only perceived at a secondary stage \cite{Russell1997}. In this sense, valence and arousal are conceptualized as ``Core Affect'' in this type of emotion theory in order to emphasize this distinction. 
Two-dimensional models of two orthogonal bipolar dimensions have been a traditional structure in dimensional models of emotion \cite{Russell1980,Russell2009,Yik2011}. Among these two-dimensional models, {\it valence} and {\it arousal} have been used very frequently ~\cite{mauss}, although variants have, e.g., suggested additional (sub-)dimensions for arousal/activation \cite{Thayer1989} or valence \cite{Watson1985}. Valence reflects the emotional sign (pleasure vs. displeasure) whereas arousal indicates a state of activation (activation vs. deactivation). While both dimensions are generally assumed to be essentially orthogonal to one another \cite{Russell1980,Russell2009,Yik2011}, there is some evidence that suggests a weak \textit{V}-shaped relationship between arousal and valence \cite{Kuppens2012}. However, the same research \cite{Kuppens2012} simultaneously highlights a large individual variation as well as the possibility of different types of context-dependent relations between both measures, thereby questioning the existence of a lawful relation between both variables. It therefore remains useful to analyze both variables separately. Coordinates in the resulting valence and arousal space can further be projected back, with some limitations, onto higher-dimensional discrete emotions models \cite{Russell1980}, e.g., fear (negative and aroused), sad (negative and not aroused) etc.. However, the use of a dimensional {\it Core Affect} structure as such does not require this translation, and has been argued to complement rather than compete with categorical structures \cite{Yik2011}. In this study, \textit{arousal} was scored using corresponding question in a questionnaire ($1\leq a\leq 7$), and positive and negative emotion subscales were merged and transformed into one value for \textit{valence} ($v=4+\frac{1}{2}(positive-negative)$; $1\leq v\leq 7$). 
There were 19 IAPS images shown to each participant. Each presentation lasted for 6 seconds and was preceded by 2 seconds of baseline and followed by an emotional questionnaire. The order of images within each IAPS set as well as the order of positive and negative trials (both in the post new thread and in the post reply sections) was varied randomly between participants. To avoid statistical effects in the physiological data that were due to fixed sequence of the stimuli, the order of both IAPS sets was furthermore randomized between participants and so was the order in which the post new thread and in the post reply sections were presented. The sequence of main experimental blocks is presented schematically in Figure 1. The sequence of events (reading of a post/ thinking about topic, contemplating the topic, writing of a post, baselines for physiological and subjective ratings, and subjective ratings of valence and arousal) within each block was always fixed and did not vary between or within participants. The whole experiment (involving forum activities) usually took about 30--40 minutes.

Main analyses were carried out using MATLAB\textsuperscript{\textregistered}\cite{matlab}, the data were stored in a PostgreSQL database. The statistical analysis of exponent comparison was performed using \textsf{R} language \cite{R}.

\section{Signals characteristics}

\begin{figure}
\centering
\begin{subfigure}[b]{0.3\textwidth}
\centering
\includegraphics[width=\textwidth]{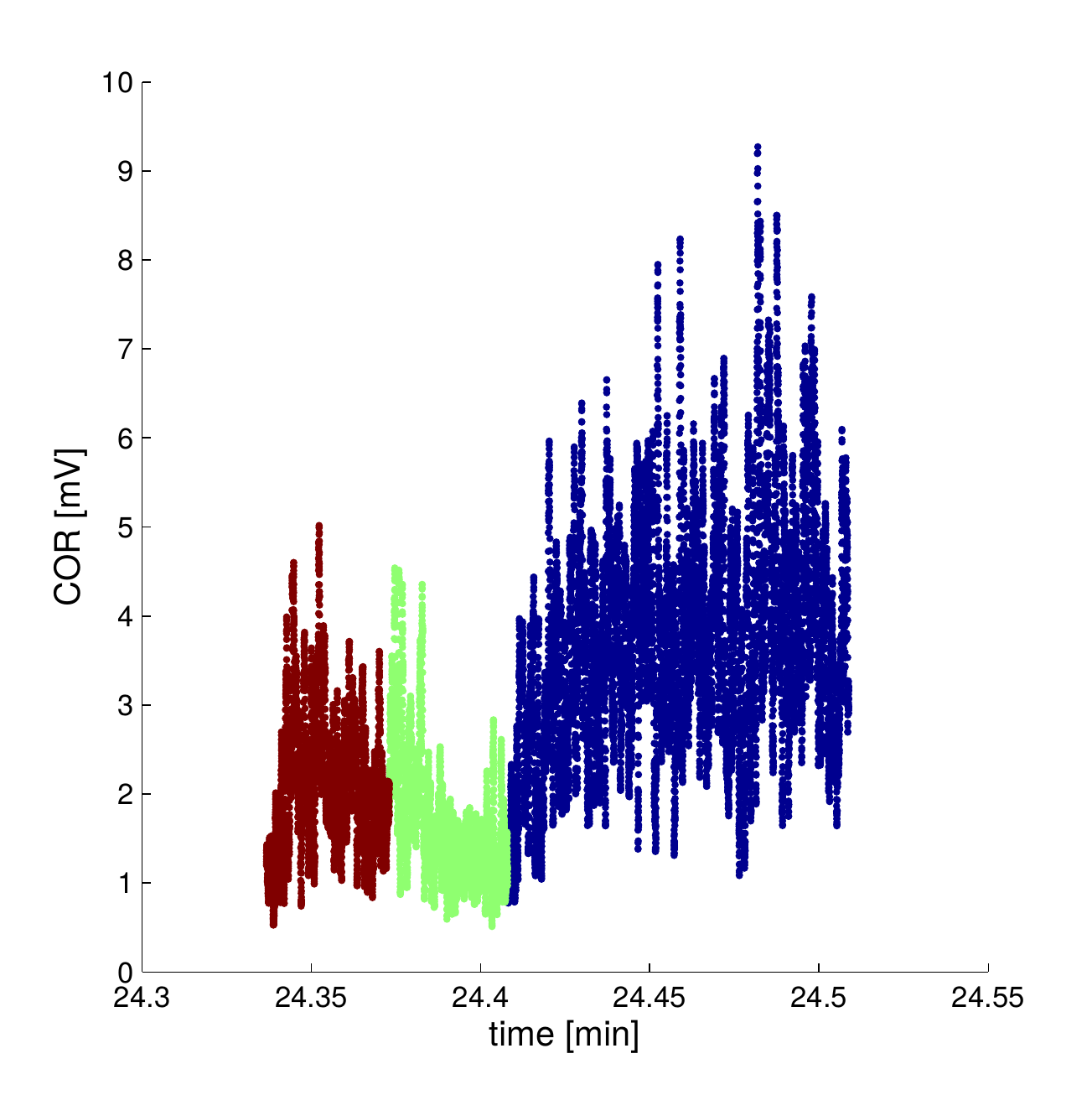}
\caption{\it COR, participant 41}
\label{fig:corpeak}
\end{subfigure}
\begin{subfigure}[b]{0.3\textwidth}
\centering
\includegraphics[width=\textwidth]{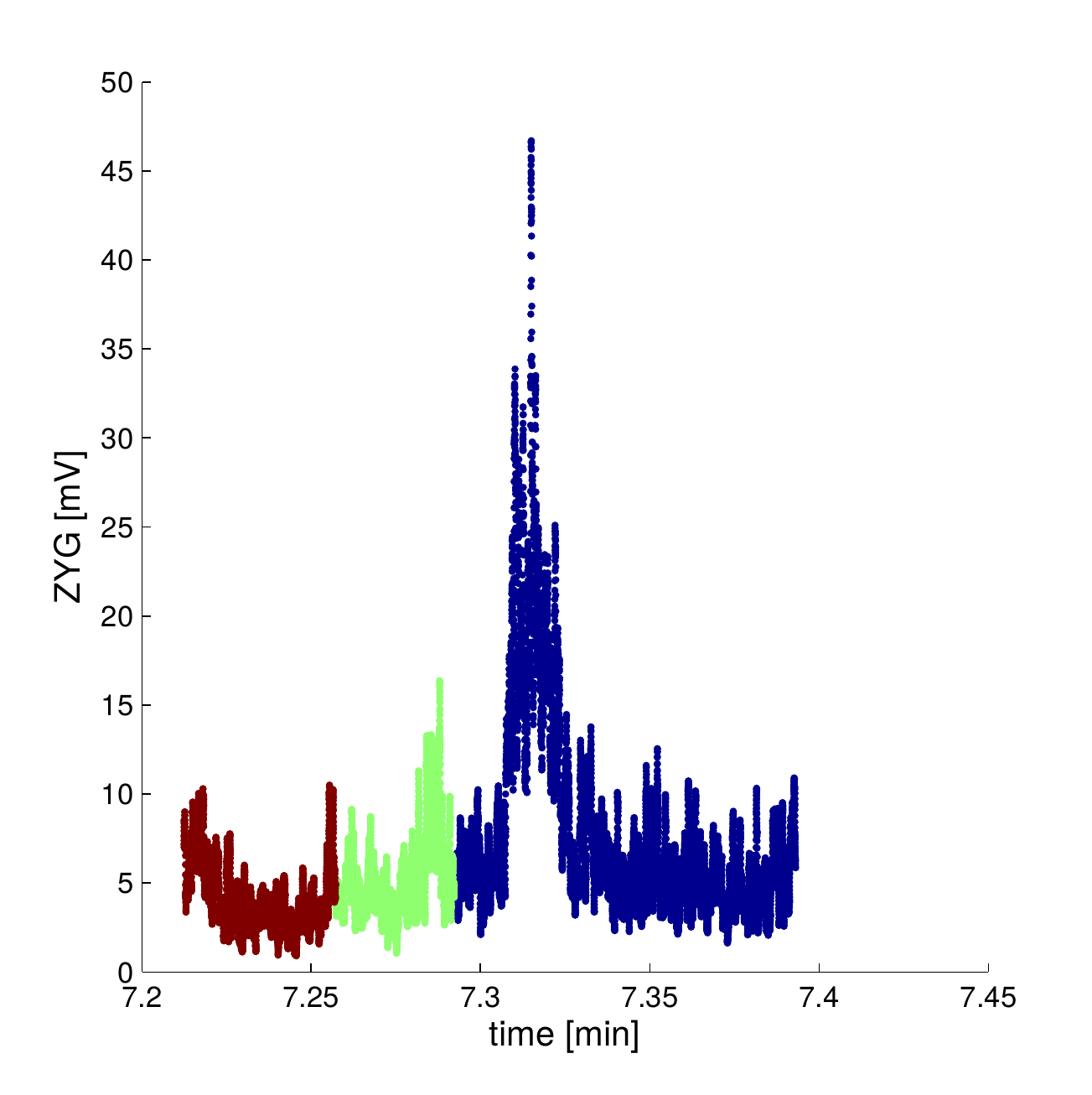}
\caption{\it ZYG, participant 46}
\label{fig:zygpeak}
\end{subfigure}
\begin{subfigure}[b]{0.3\textwidth}
\centering
\includegraphics[width=\textwidth]{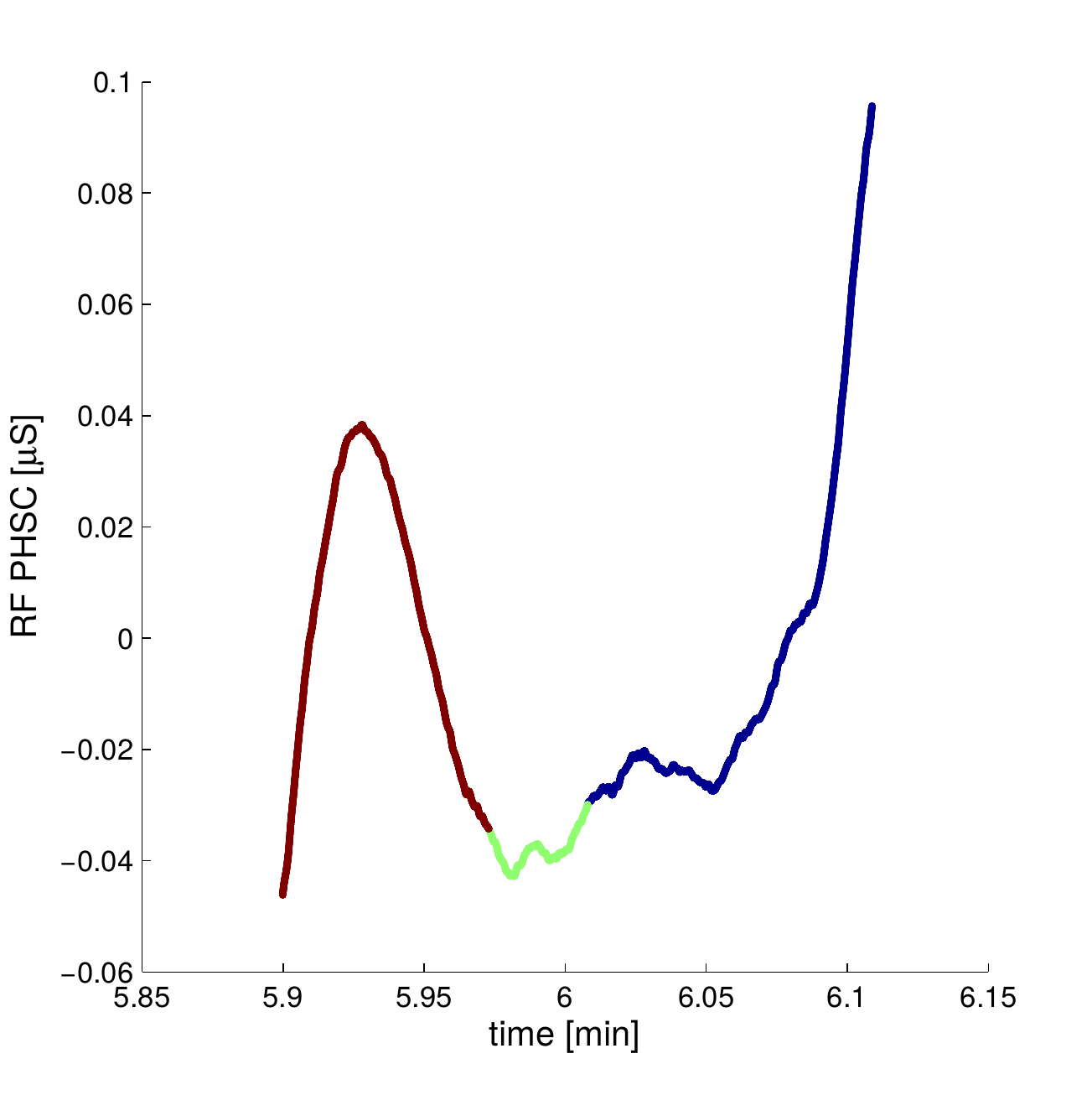}
\caption{\it PHSC, participant 61}
\label{fig:phscpeak}
\end{subfigure}
\caption{Example of signals' responses to an emotional stimulus (IAPS 6260); colors indicate current event -- brown - introduction to IAPS demonstration, green - baseline, dark blue - stimulus presentation.}\label{fig:peaks}
\end{figure}

\subsection{Facial electromyography}
\par Facial electromyography (fEMG) is a well established method for the measurement of facial activity, including facial muscle activity associated with emotional valence \cite{tassinary}. In the present study, this measurement focused on activation over two sites of facial muscles. \textit{Corrugator supercilii} (COR; exemplary response to a stimulus - see~\ref{fig:corpeak}) muscle activity is exhibited when frowning, and shows a negative linear correlation with valence of experienced emotion -- less activity in response toward more pleasant stimuli; activity at the \textit{zygomaticus major} (ZYG; exemplary response to a stimulus - see~\ref{fig:zygpeak}) muscle site is associated with smiling, and a quadratic effect of valence (i.e. highest activities are obtained for extreme emotions) \cite{larsen}. One cannot exactly map the activity of the muscles with corresponding emotions or even facial expressions, because of the variety of uses of these muscles (e.g., during speech), as well as their role in social interactions including, e.g., polite smiling that does not express an intense internal emotional state as such. However, in the conditions of controlled laboratory experiments such as the present research, facial activity unrelated to emotions is occasional and can be regarded as error variance across comparable conditions of emotional stimulation.
\par All analyses were performed using raw signals because any smoothing or filtering would cause a loss of information about signal fluctuations.
\par The COR and ZYG signals are somewhat similar due to their origin, namely muscle activity. They differ in that COR is bilaterally innervated as opposed to a greater contralateral innervation of ZYG \cite{Rinn}.

\subsection{Electrodermal activity}
\par Electrodermal activity (EDA), or skin conductance (SC) analyses are based on Galvanic skin responses, i.e., changes in the electrical conductivity of the skin that are most typically recorded at the subject's hands (palmar) or feet (plantar) \cite{boucsein}. As has been known already since the late 1920s \cite{Darrow1927,dawson}, these changes are related to the opening and closing of sweat glands in the skin that produce sweating, which in turn is known to be related to arousal \cite{alex,dawson,boucsein}. These changes in phasic EDA can be caused by experiencing an emotion (such as being exposed to various visual stimuli) \cite{bach2}. More specifically, both \textit{tonic} changes in \textit{skin conductance level} (SCL) as well as \textit{phasic} \textit{skin conductance responses} (SCRs) have frequently been used in the literature as indicators of sympathetic emotional arousal \cite{boucsein}. The \textit{tonic} part of the signal reflects a slowly changing global trend that is not directly in response to short-term visual stimulation, and therefore was not used for the present fluctuation-dissipation analyses. The \textit{phasic} part is a so-called rapid changing factor. SCRs are a type of phasic response that are widely used in scoring event-related arousal exhibited by experimental subjects. SCRs are typically defined within the psychophysiological literature \cite{dawson} as requiring a certain minimum amplitude of peak such as 0.01-0.05 $\mu$S (microsiemenses -- unit of electric conductance). In addition, \textit{event-related SCRs} must occur within a specific time window (e.g., 1-4 s latency) in order to qualify as likely related to an external event, whereas non-specific SCRs (NS-SCRs) can occur at any time during the recording. SCRs will subsequently be referred to simply as \textit{phasic SC} (PHSC; exemplary response to a stimulus - see~\ref{fig:phscpeak}). In the study, we analyzed only 6 seconds intervals of stimuli presentation thus we assume that all observed peaks were event related.

\par The present research involved a bilateral plantar recording of EDA, i.e., from the subjects' feet. As opposed to so-called non-palmar non-plantar sites \cite{boucsein}, an adequate recording at this site is non-controversial since it has been shown to exhibit good measurement properties and is to be preferred over recordings from, e.g., the wrist, which is more affected by thermoregulation \cite{boucsein}. Palmar recording sites (i.e., at the palms or fingers) are even more typical for laboratory measurements of EDA. However, in the present study, participants had to type on a keyboard during the experiment. This would have resulted in substantial movement artifacts for a palmar measure. The plantar recording sites greatly minimized this issue, and further allowed a bilateral recording, allowing a validation of the recording quality in the event of any remaining movement artefacts. However, no significant intraindividual differences between left and right foot EDA signals were observed (mean Pearson correlation coefficient = 0.97; standard deviation in the population = 0.02). Unless stated otherwise, the right foot SCL signal was analyzed.

\section{Results}

We are interested in quantifying temporal fluctuation scaling (TFS) for time series obtained by facial electromyography (two sets: aggregated activity of COR and ZYG muscles separately) and electrodermal activity (one set: aggregated effect from sweat glands --- PHSC). Our first analysis concerns the {\it whole} signal, i.e., a complete time series acquired during the experiment that include, inter alia, episodes of emotional simulations with IAPS images. In the second stage we try to use additional data to separate series connected to specific levels of emotions (valence / arousal levels). The third and last set of results contains an analysis of the windows size influence on the obtained scaling exponents.
\begin{figure}[!bh]
\centering
\begin{subfigure}[b]{0.32\textwidth}
\centering
\includegraphics[width=\textwidth]{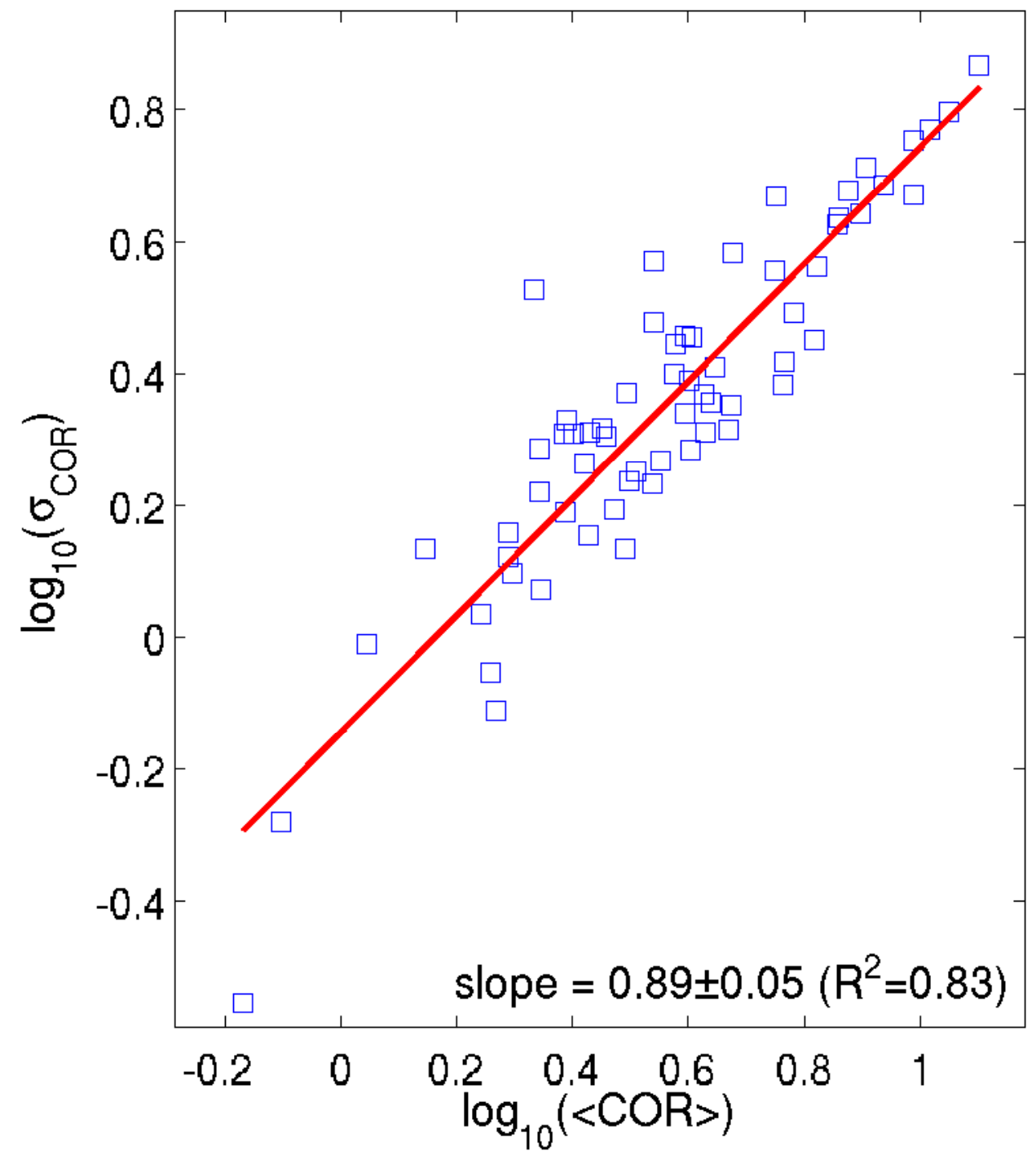}
\caption{\it COR}
\label{fig:tfs_cor}
\end{subfigure}
\begin{subfigure}[b]{0.32\textwidth}
\centering
\includegraphics[width=\textwidth]{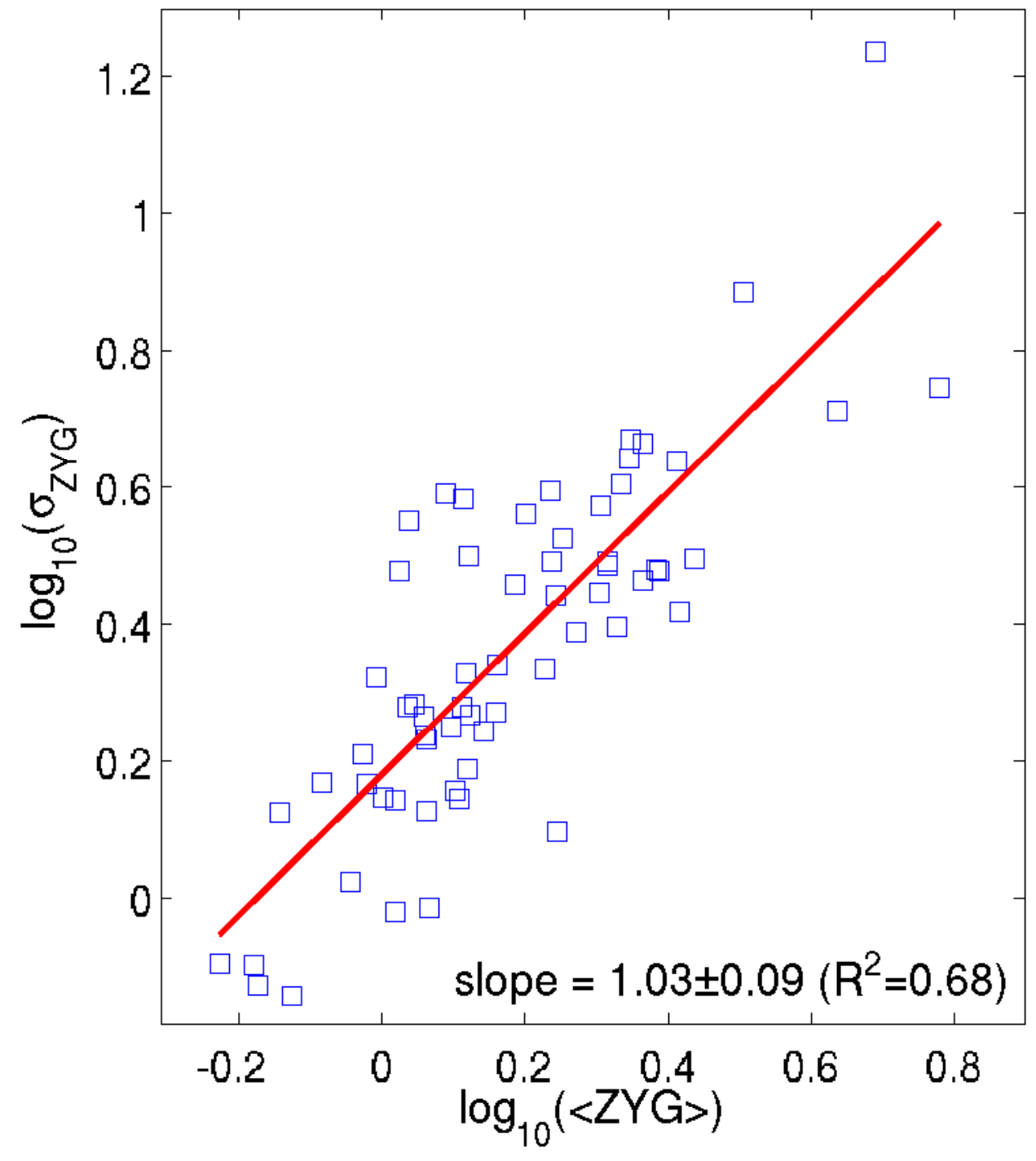}
\caption{\it ZYG}
\label{fig:tfs_zyg}
\end{subfigure}
\begin{subfigure}[b]{0.32\textwidth}
\centering
\includegraphics[width=\textwidth]{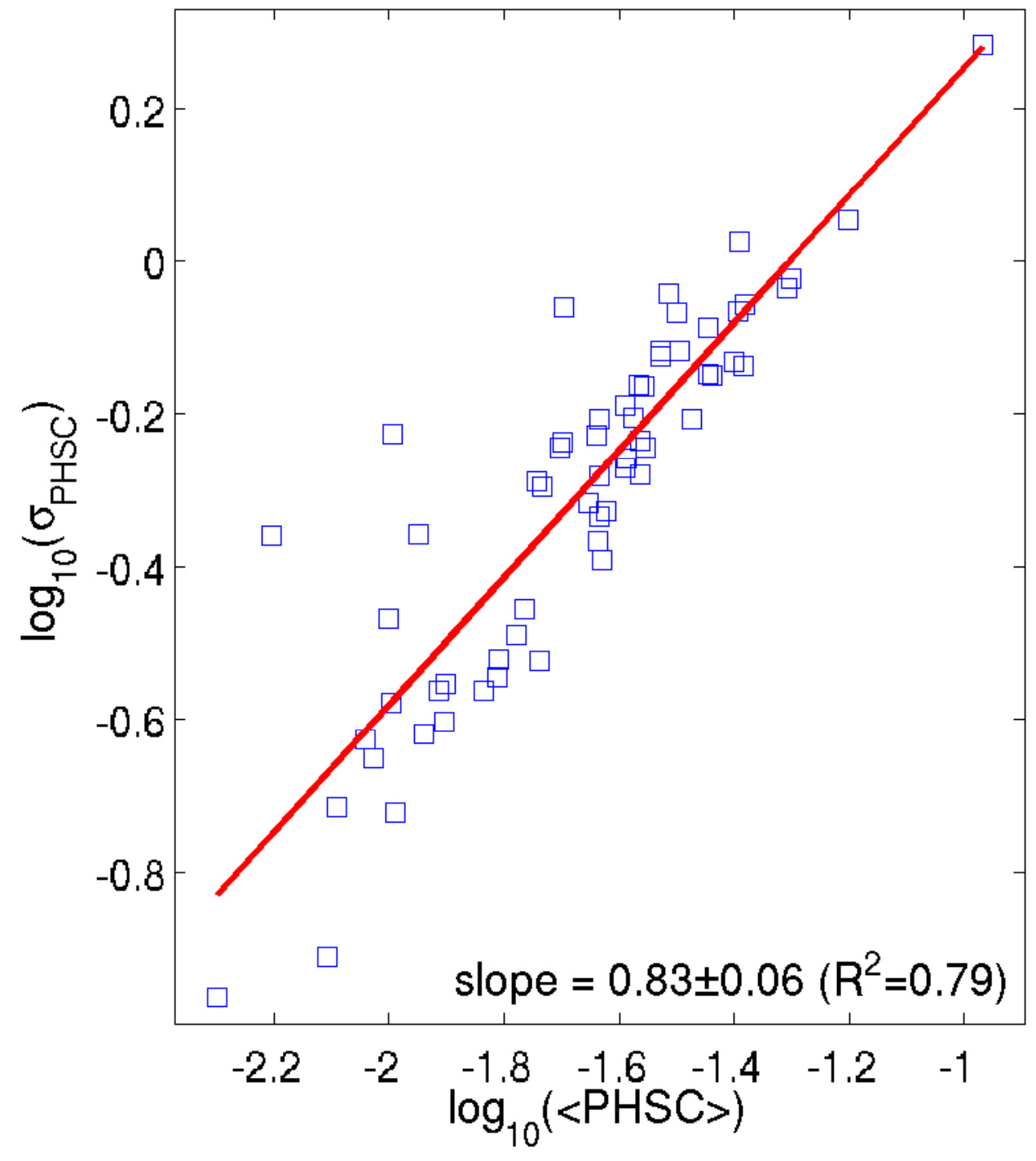}
\caption{\it PHSC}
\label{fig:tfs_phsc}
\end{subfigure}
\caption{\textbf{Temporal fluctuation scaling} for \textbf{the whole signals} with $\Delta = 1/\nu_s=0.0005\mathrm{s}$. Standard deviation as a function of the mean values of signals; each point represents a signal of one participant; least squares linear regression was applied to log-log data ($y=ax^b \Rightarrow \log y = \log a + b \log x$); errors are standard errors obtained using the least squares method.}
\label{fig:tfs-all}
\end{figure}
\subsection{Taylor's scaling}
Figure \ref{fig:tfs-all} presents a scatter plot of standard deviation versus mean characteristics for all the participants and different signals --- COR (Fig. \ref{fig:tfs-all}a), ZYG (Fig. \ref{fig:tfs-all}b) and PHSC (Fig. \ref{fig:tfs-all}c). When considering a time series for the whole experiment, one has to realize that the each participant experiences different kinds and magnitudes of stimulation. It follows that the measured variances originate not only from fast fluctuations of muscle activity or skin conductance but also from passing through several emotional states induced by different types of IAPS images. Nonetheless, for all the mentioned signals we are able to observe Taylor's scaling characterized by exponents (values shown with standard error), respectively $\alpha_{COR}=0.89 \pm 0.05$ (goodness of fit: $R^2=0.83$), $\alpha_{ZYG}=1.03 \pm 0.09$ ($R^2=0.68$) and $\alpha_{PHSC}=0.86\pm 0.07$ ($R^2=0.86$) that are significantly higher than $\alpha = 0.5$. Such results might suggest a partial synchronization of elements which contribute to the value of the signal.
\subsection{Separation of specific emotions}
In addition to raw time series the dataset contains also information about participants' answers included in the emotional questionnaires. We further assume that those answers reflect emotional state of participant, and that they are comparable between participants (the same scores given by different participants describe similar emotion). Owing to the set of markers that had been placed in the time series (see Sec. \ref{sec:data}) we were able to divide the signals connected to each participant and IAPS image into groups of short (6 seconds --- time of IAPS exposition) time subseries related to different emotional states. Such a setting gives us an opportunity to check if the fluctuation scaling exponents characterize each of the states and if they can be distinguished basing on them. In order to conduct this analysis one also needs to address the choice of the time window size $\Delta$. The plots shown in Fig.~\ref{fig:tfs-all} were obtained using the smallest possible time window, i.e., $\Delta = \nu_s^{-1}=0.0005\mathrm{s}$, where $\nu_s =$ 2 kHz is the sampling frequency. Knowing that the size of time window might influence the results, an analysis with aggregation for each IAPS--participant pair with a given score was performed using different $\Delta$ in order to find one which should allow to distinguish between emotions. The values of $\Delta$ ranging from $5$ ms (minimum of 10 samples per window) to $1.5$s (minimum of 4 full windows per IAPS-participant pair) were considered.

\begin{figure}[t]
\centering
\begin{subfigure}[b]{0.48\textwidth}
\centering
\includegraphics[width=\textwidth]{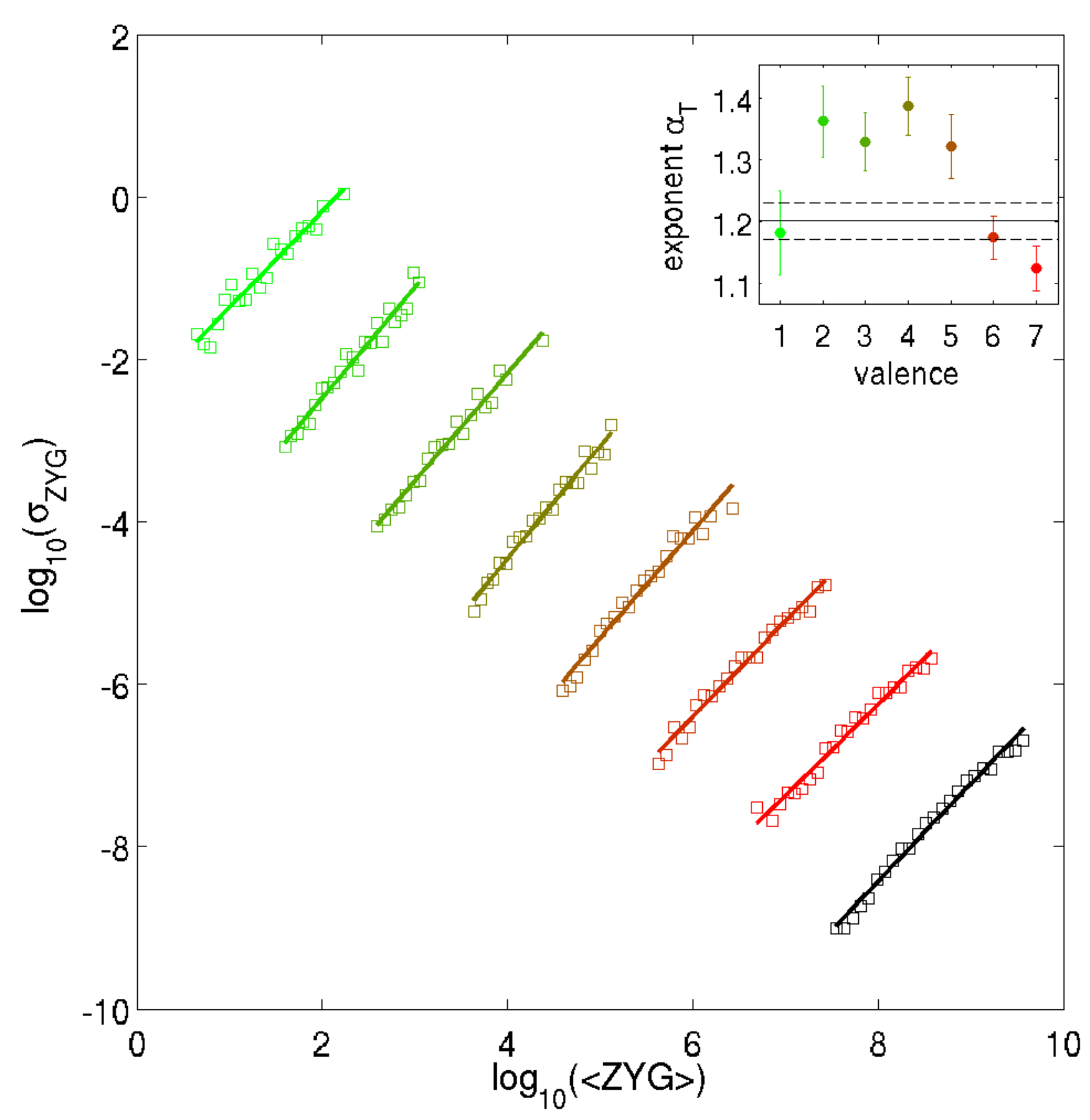}
\caption{\it ZYG (valence)}
\label{fig:tfs-iaps-zyg-val}
\end{subfigure}
\begin{subfigure}[b]{0.48\textwidth}
\centering
\includegraphics[width=\textwidth]{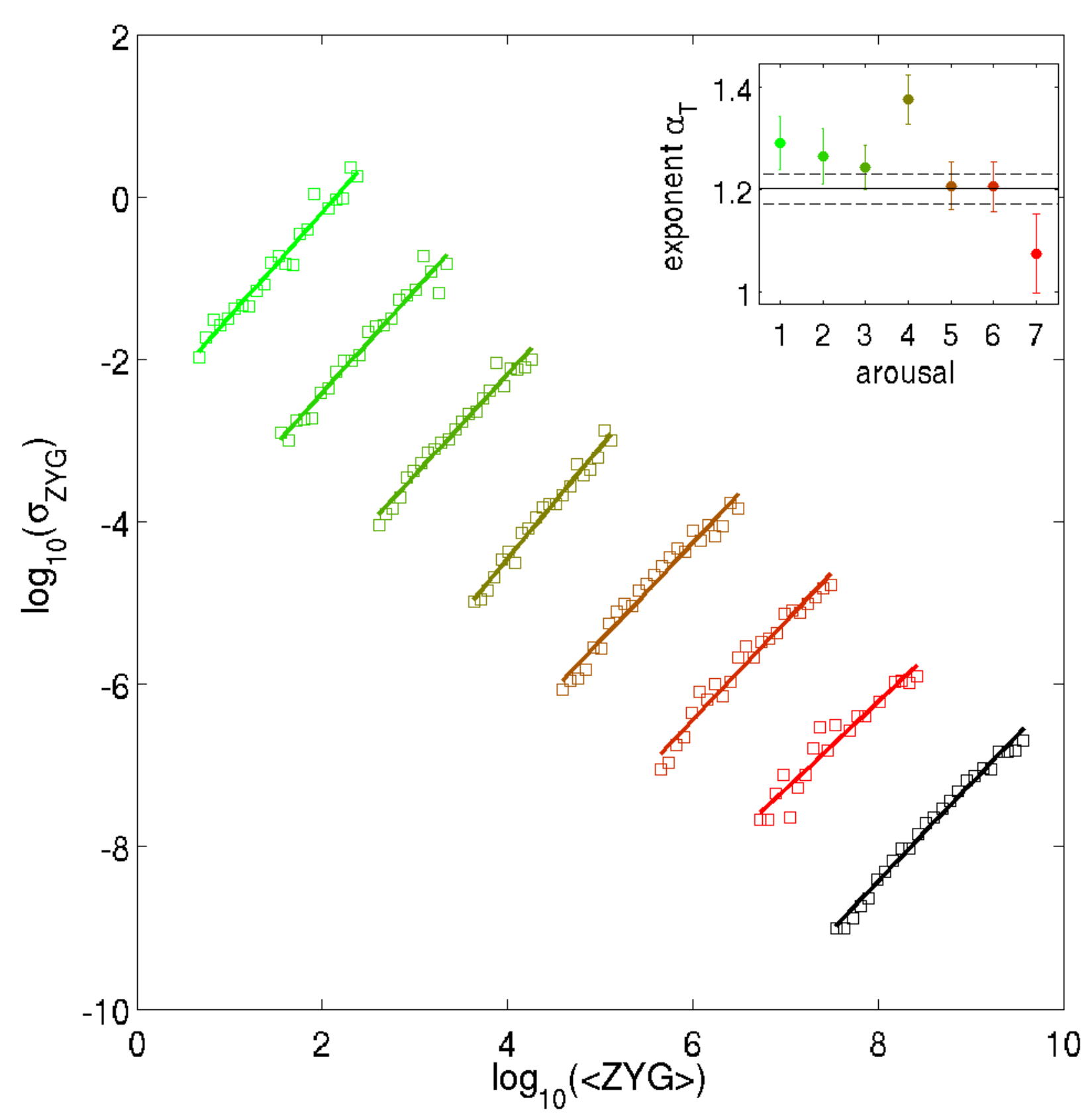}
\caption{\it ZYG (arousal)}
\label{fig:tfs-iaps-zyg-aro}
\end{subfigure}
\caption{Results of temporal fluctuation measured for \textbf{ZYG} by standard deviations as functions of mean values of this signal during \textbf{IAPS stimulation}. Different colors correspond to   specific values  of IAPS emotional (left) \textbf{valence} score, from very negative emotions ($v=1$ - light green) to very positive ($v=7$, dark red) ones, (right) \textbf{arousal} score, from very calm ($a=1$ - light green) to very aroused ($a=7$, dark red) states. The black line -- all IAPS data aggregated. Results are binned logarithmically in 25 bins (each point is a logarithm of mean value of $\sigma_{ZYG}$ in all time windows with $\log_{10}\langle ZYG\rangle$ in a given bin) and shifted both vertically and horizontally for the sake of readability; Insets show values of  exponents $\alpha$, i.e slope of the plotted lines  as a function of a given score (black horizontal  line -- exponent for all IAPS data aggregated, dotted black lines $\alpha\pm\sigma(\alpha)$.). Observation window   $\Delta =0.04\mathrm{s}$ was used.}
\label{fig:tfs-iaps-zyg}
\end{figure}

\begin{figure}[t]
\centering
\begin{subfigure}[b]{0.48\textwidth}
\centering
\includegraphics[width=\textwidth]{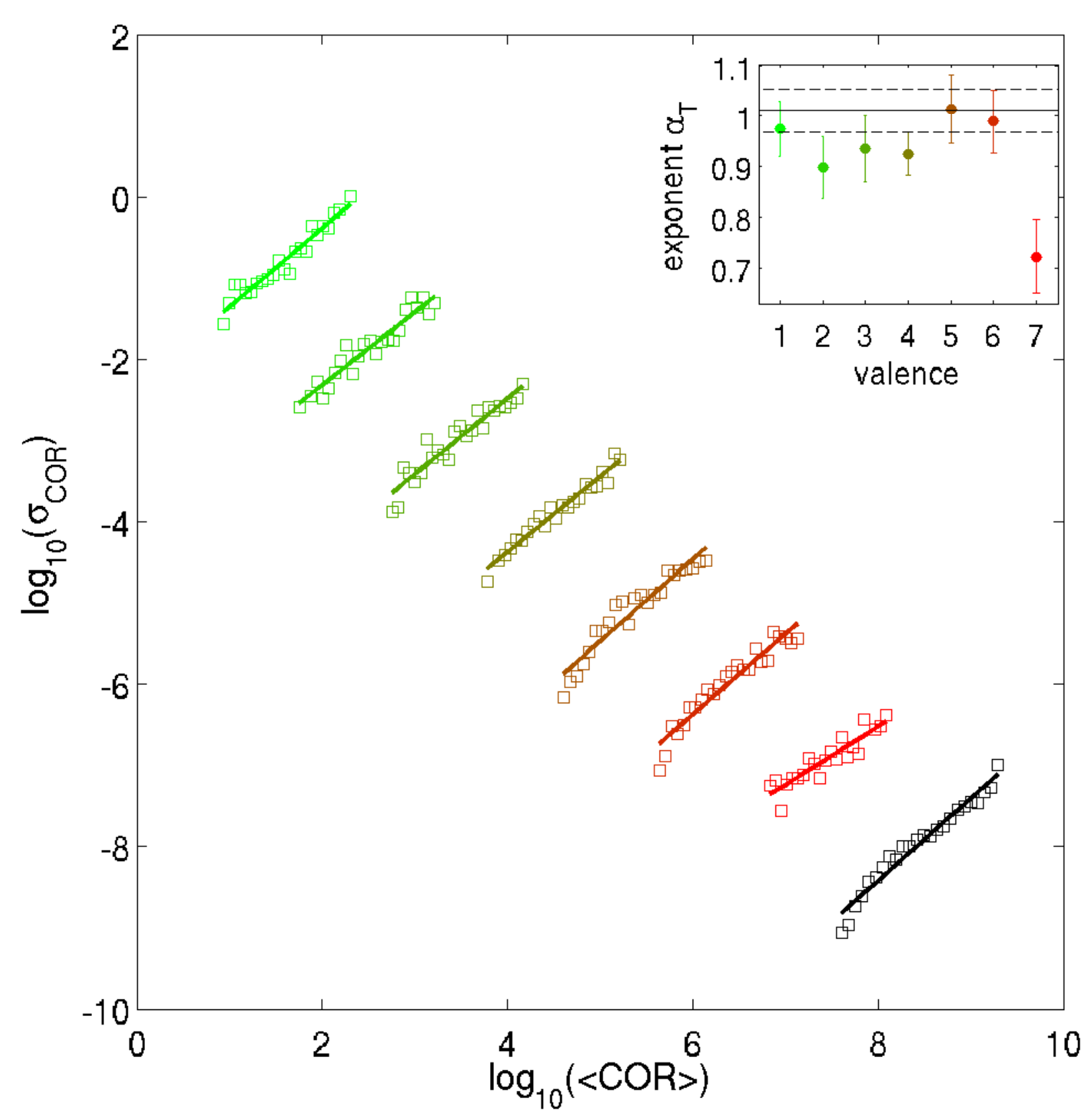}
\caption{\it COR (valence)}
\label{fig:tfs-iaps-cor-val}
\end{subfigure}
\begin{subfigure}[b]{0.48\textwidth}
\centering
\includegraphics[width=\textwidth]{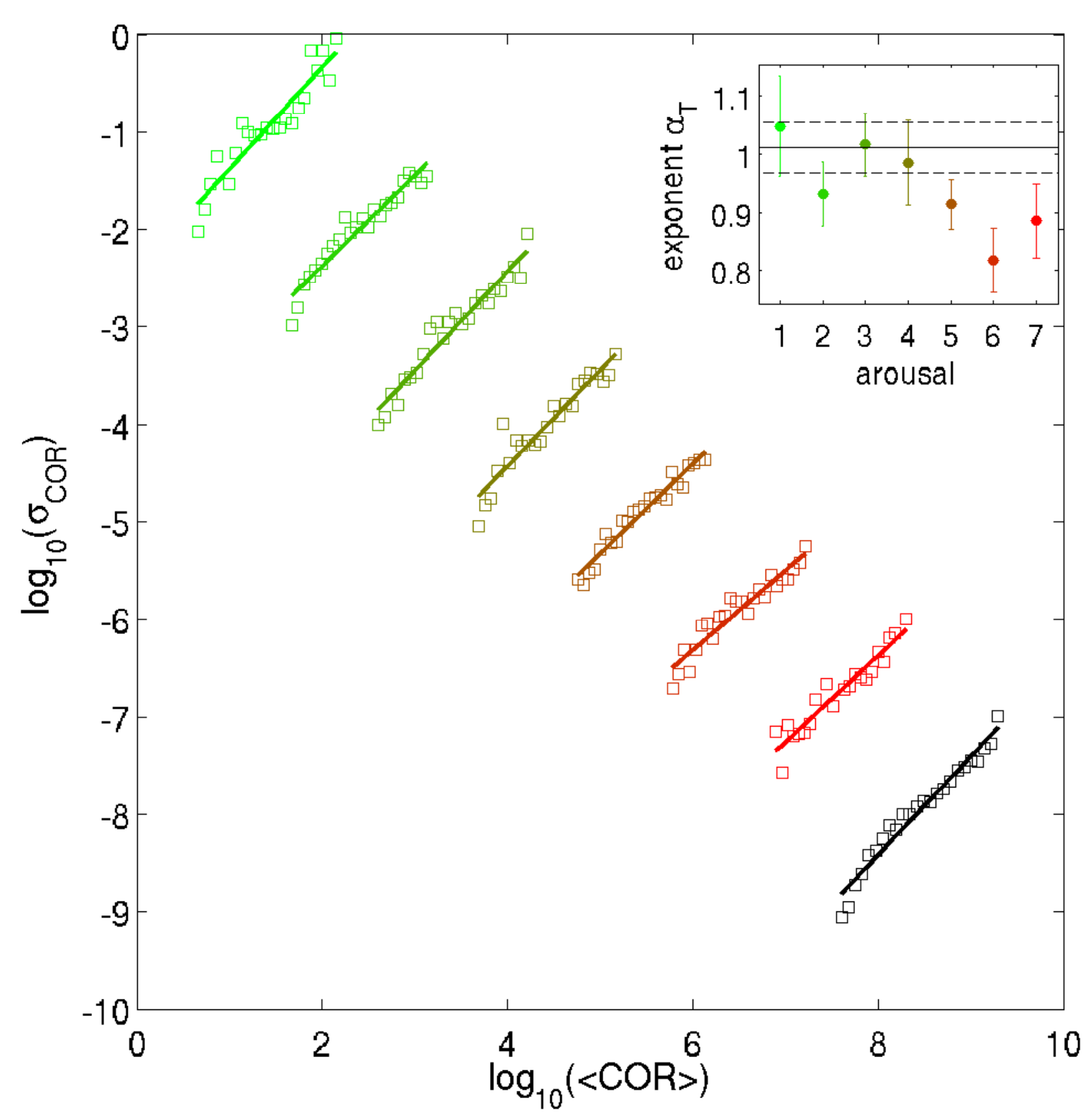}
\caption{\it COR (arousal)}
\label{fig:tfs-iaps-cor-aro}
\end{subfigure}
\caption{Results of temporal fluctuation measured for \textbf{COR} by standard deviations as functions of mean values of this signal during \textbf{IAPS stimulation}. Different colors correspond to   specific values  of IAPS emotional (left) \textbf{valence} score, from very negative emotions ($v=1$ - light green) to very positive ($v=7$, dark red) ones, (right) \textbf{arousal} score, from very calm ($a=1$ - light green) to very aroused ($a=7$, dark red) states. The black line -- all IAPS data aggregated. Results are binned logarithmically in 25 bins (each point is a logarithm of mean value of $\sigma_{COR}$ in all time windows with $\log_{10}\langle COR\rangle$ in a given bin) and shifted both vertically and horizontally for the sake of readability; Insets show values of  exponents $\alpha$, i.e slope of the plotted lines  as a function of a given score (black horizontal line -- exponent for all IAPS data aggregated, dotted black lines $\alpha\pm\sigma(\alpha)$.). Observation window   $\Delta =0.36\mathrm{s}$ was used. }
\label{fig:tfs-iaps-cor}
\end{figure}

\begin{figure}[t]
\centering
\includegraphics[width=.48\textwidth]{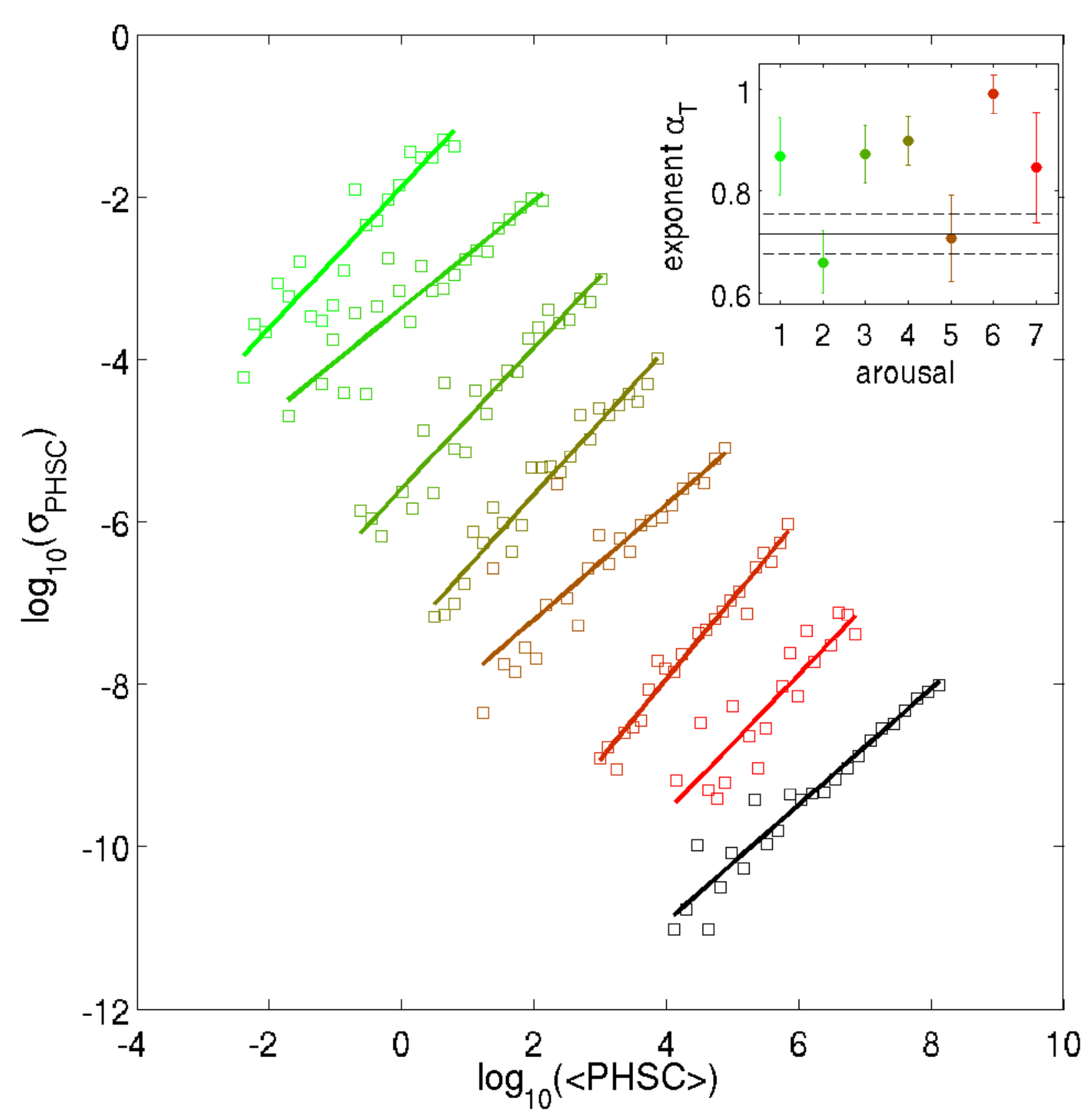}
\caption{Results of temporal fluctuation measured for \textbf{PHSC} by standard deviations as functions of mean values of this signal during \textbf{IAPS stimulation}. Different colors correspond to   specific values  of IAPS emotional \textbf{arousal} score (for valence no significant differences were found), from very calm ($a=1$ - light green) to very aroused ($a=7$, dark red) states. The black line -- all IAPS data aggregated. Results are binned logarithmically in 25 bins (each point is a logarithm of mean value of $\sigma_{PHSC}$ in all time windows with $\log_{10}\langle PHSC\rangle$ in a given bin) and shifted both vertically and horizontally for the sake of readability; Insets show values of  exponents $\alpha$, i.e slope of the plotted lines  as a function of a given score (black horizontal line -- exponent for all IAPS data aggregated, dotted black lines $\alpha\pm\sigma(\alpha)$. The observation window $\Delta =0.04\mathrm{s}$ was used.}
\label{fig:tfs-iaps-phsc}
\end{figure}

To address the above presented issues we performed analysis of covariance (ANCOVA) for each signal and each observation window size $\Delta$. In this way we are able to check if the interaction term (average $\times$ level of valence / arousal) is statistically significant, which in turn allows us to test the hypothesis of equal slopes ($\alpha$ exponents) for different values of valence/arousal. The results of the analysis are presented in Table \ref{stat:ancova} leading to two instant conclusions for valence levels: (i) there is a strong difference among the scaling exponents for ZYG signal and (ii) there is no statistical difference among the scaling exponents in PHSC signal. For COR signal we observe a mixed effect. In the case of arousal the interpretation is also far from being straightforward: arousal levels seem to distinguishable for the majority of $\Delta$ windows in PHSC signal, while in case of COR the difference is seen only for large $\Delta$. Opposite to that there is some mild statistical evidence that arousal levels can be distinguished for small $\Delta$ in ZYG signal. The window sizes which yield the highest significance ($\Delta=0.04\mathrm{s}$ for ZYG and PHSC, $\Delta = 0.36\mathrm{s}$ for COR) were used in the next part of the analysis. For those windows a mean value of a given signal was obtained and then the mean and standard deviations of those mean values were calculated for each IAPS-participant pair. For pairs with a given emotional score, results were divided into 25 logarithmic bins and a linear regression using a least squares method was performed. Results of the fluctuations scaling analysis for each signal and emotional score is presented in Figs.~\ref{fig:tfs-iaps-zyg}---\ref{fig:tfs-iaps-phsc}. Insets present values of scaling exponents $\alpha$ as a function of valence/arousal levels.

\begin{table}[!ht]
\centering
\begin{tabular}{lccccccccc}
  \hline
signal / $\Delta$[s] & 0.005 & 0.01 & 0.02 & 0.04 & 0.09 & 0.18 & 0.36 & 0.74 & 1.51 \\ 
  \hline
ZYG $v$ & $^{***}$ & $^{***}$ & $^{***}$ & $^{***}$ & $^{***}$ & $^{***}$ & $^{**}$ & $^{***}$ & $^{***}$ \\ 
  ZYG $a$ & . & $^{*}$ & $^{*}$ & $^{*}$ & . & . &  &  &  \\ 
  COR $v$ &  &  &  & . &  & $^{*}$ & $^{*}$ & . & . \\ 
  COR $a$ &  &  &  &  &  &  & $^{*}$ & $^{*}$ & $^{**}$ \\ 
  PHSC $v$ &  &  &  &  &  &  &  &  &  \\ 
  PHSC $a$ & $^{*}$ & $^{**}$ & $^{*}$ & $^{**}$ &  & $^{*}$ & . &  & $^{**}$ \\ 
   \hline
\end{tabular}
\caption{{\bf Results of the analysis of covariance (ANCOVA)} for different signals and different time windows $\Delta$. The following significance codes to express p-values $p$ of the ANCOVA $F$-test are used: $^{***}$ for $p < 0.001$, $^{**}$ for $0.001 < p < 0.01$, $^{*}$ for $0.01 < p < 0.05$, . for $0.05 < p < 0.1$ and empty space for $p > 0.1$.}  
\label{stat:ancova}
\end{table}
Let us now briefly inspect those insets to describe the results in qualitative way. Surprisingly for ZYG signal we have $\alpha>1$ for all valence and arousal levels. Moreover for extreme values of emotional valencies ($v=1,6,7$; Fig.~\ref{fig:tfs-iaps-zyg-val}) $\alpha$ reaches its lowest values that seem to be different than those for other scores. Additionally highly aroused state ($a=7$; Fig.~\ref{fig:tfs-iaps-zyg-aro}) can be separated from all other scores with its exponent being the lowest one. In COR signal, scaling exponents for the majority of the series show $\alpha\approx1$ but they drop to $\alpha\approx 0.85$ for very positive ($v=7$; Fig.~\ref{fig:tfs-iaps-cor-val}) and very aroused states ($a=6,7$; Fig.~\ref{fig:tfs-iaps-cor-aro}). Finally the results in the case of PHSC signal for arousal (Fig.~\ref{fig:tfs-iaps-phsc}) are very noisy and no clear trends are visible except for outlying character of $a=2,5,6$ levels. There is no figure for valence as results were statistically insignificant (see below). The assumption of non-negativity of the signal is not fulfilled --- some results had negative means (which is an artifact of separating measured signal to \textit{phasic} and \textit{tonic} parts) and thus they were discarded.

In order to statistically infer differences between specific exponents, we treat valence and arousal levels as dummy variables in the regression analysis and compare coefficient by $t$-tests treating consecutive levels (i.e., $v = 1$, $v=2$, etc.) as reference values. We use false discovery rate \cite{bh1995} to adjust originally obtained p-values controlling the expected proportion of false discoveries among the rejected hypotheses. The exponents' comparison for selected $\Delta$ ($\Delta$ giving the highest significance in Table \ref{stat:ancova}) is shown in Tables \ref{stat:zyg}-\ref{stat:phsc}, where we use significance codes to express p-values that allow for an instant inspection of the differences among the results.   

Statistical analyses back up our previous conclusions. Indeed, in the case of valence in ZYG (see Table \ref{stat:zyg}-left), the extreme cases ($v=1,6,7$) are different from the rest ($v = 2,3,4,5$), which in turn are indistinguishable. On the other hand, the differences among $v=1,6,7$ levels are not significant which might suggest a parabola-like relation between $\alpha$ exponent and valence $v$. In the case of arousal (see Table \ref{stat:zyg}-right) only the most aroused state ($a=7$) differs from $a=1$ and $a=4$. For valence in COR signal (Table \ref{stat:cor}-left), the situation is even more obvious --- only $v=7$ is significantly different in this set, other points seem to form a stable level (with a single exception of $v=2$). The case of arousal in COR (Table \ref{stat:cor}-right) resembles the same variable in ZYG: here $a=6$ differs from $a=1$ and $a=3$. We do not perform exponents comparison for valence in PHSC because of the lack of significance in the analysis of covariance (see Table \ref{stat:ancova}). Finally, for arousal in this signal (Table \ref{stat:phsc}) we have two levels ($a = 2,6$) that are significantly different from the majority of other scores.

\subsection{Window size analysis}
\begin{figure}[tb]
\centering
\begin{subfigure}[b]{0.45\textwidth}
\centering
\includegraphics[width=\textwidth]{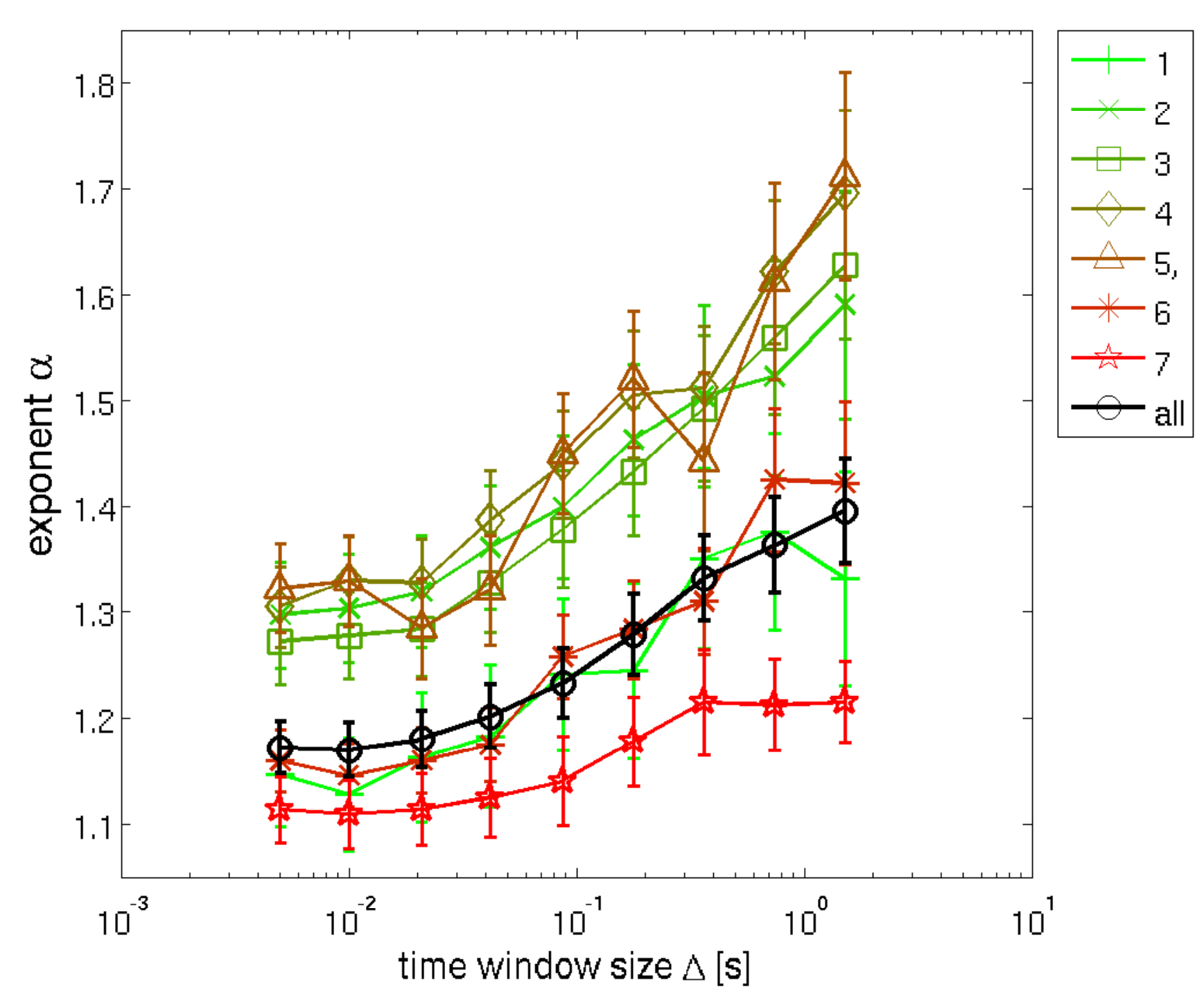}
\caption{\it ZYG (valence)}
\label{fig:alfa-zyg-val}
\end{subfigure}
\begin{subfigure}[b]{0.45\textwidth}
\centering
\includegraphics[width=\textwidth]{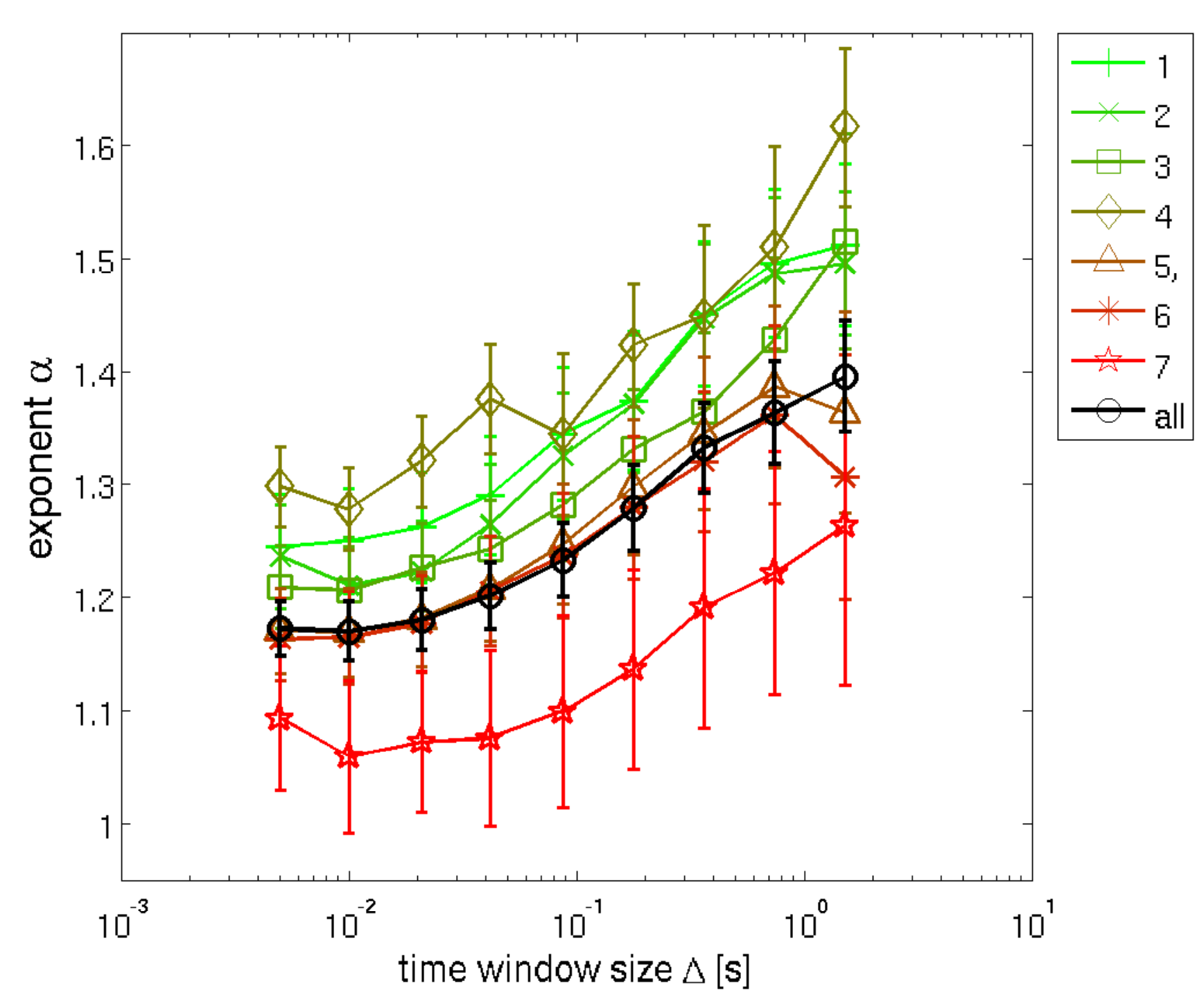}
\caption{\it ZYG (arousal)}
\label{fig:alfa-zyg-aro}
\end{subfigure}
\begin{subfigure}[b]{0.45\textwidth}
\centering
\includegraphics[width=\textwidth]{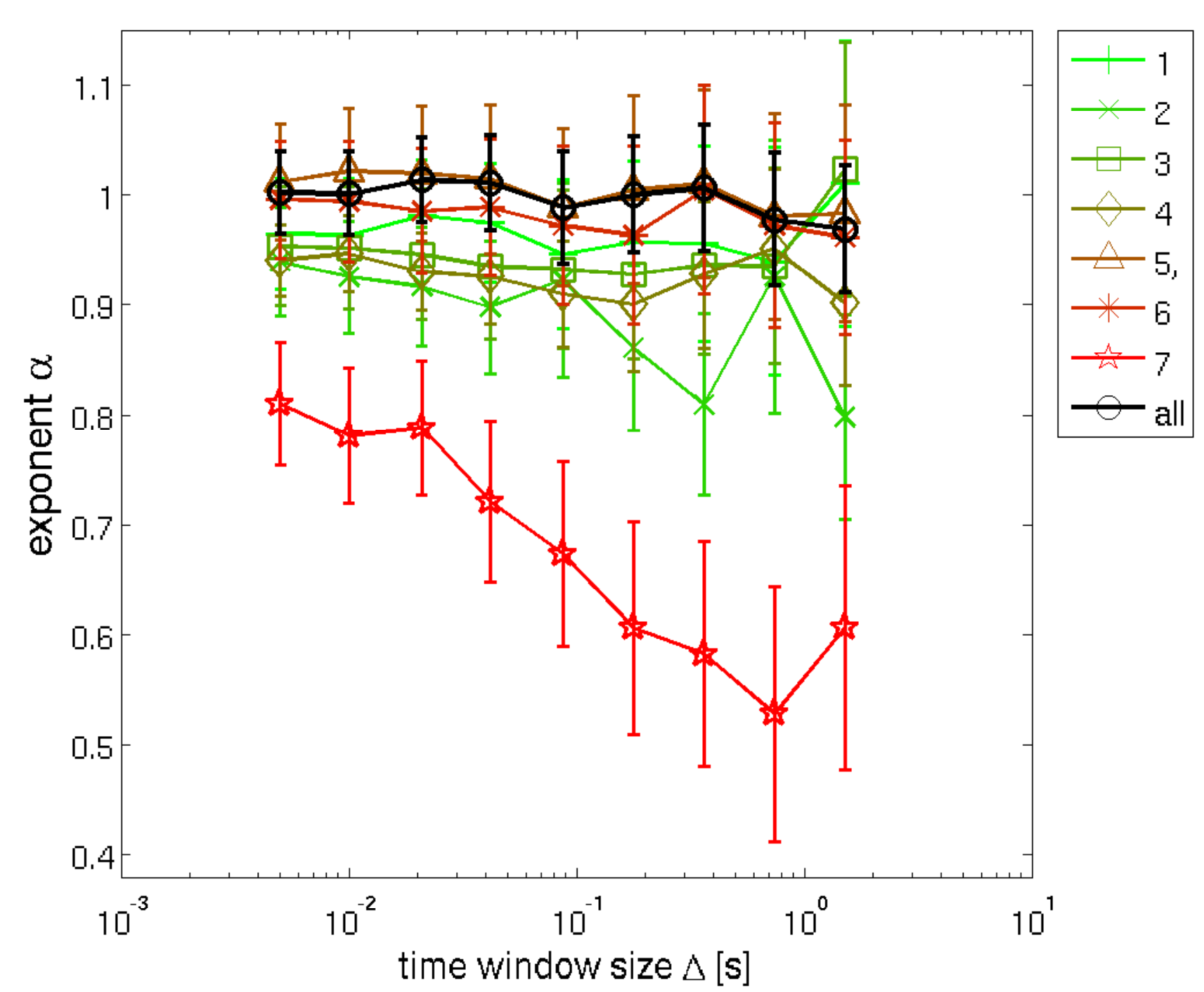}
\caption{\it COR (valence)}
\label{fig:alfa-cor-val}
\end{subfigure}
\begin{subfigure}[b]{0.45\textwidth}
\centering
\includegraphics[width=\textwidth]{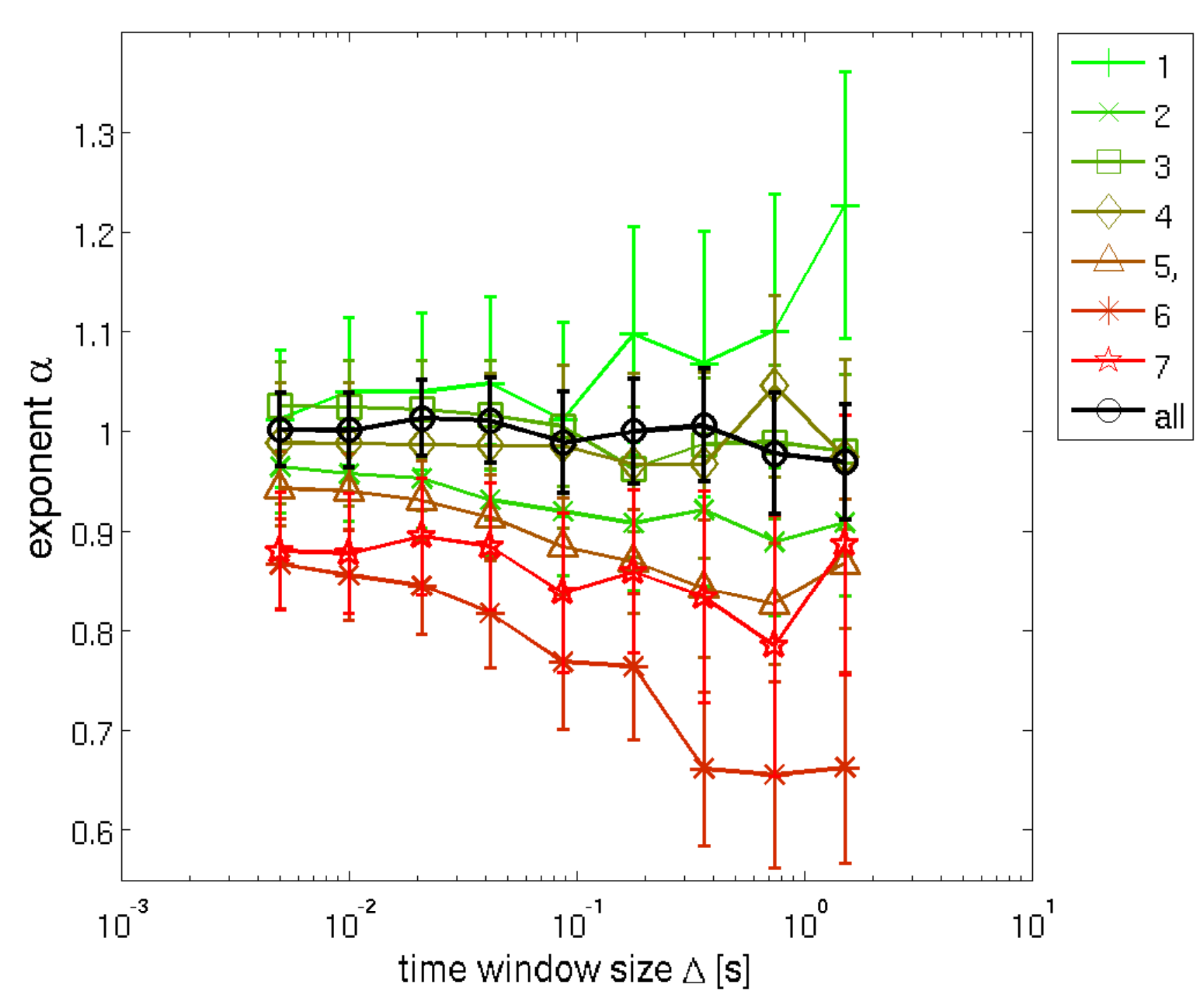}
\caption{\it COR (arousal)}
\label{fig:alfa-cor-aro}
\end{subfigure}
\begin{subfigure}[b]{0.45\textwidth}
\centering
\includegraphics[width=\textwidth]{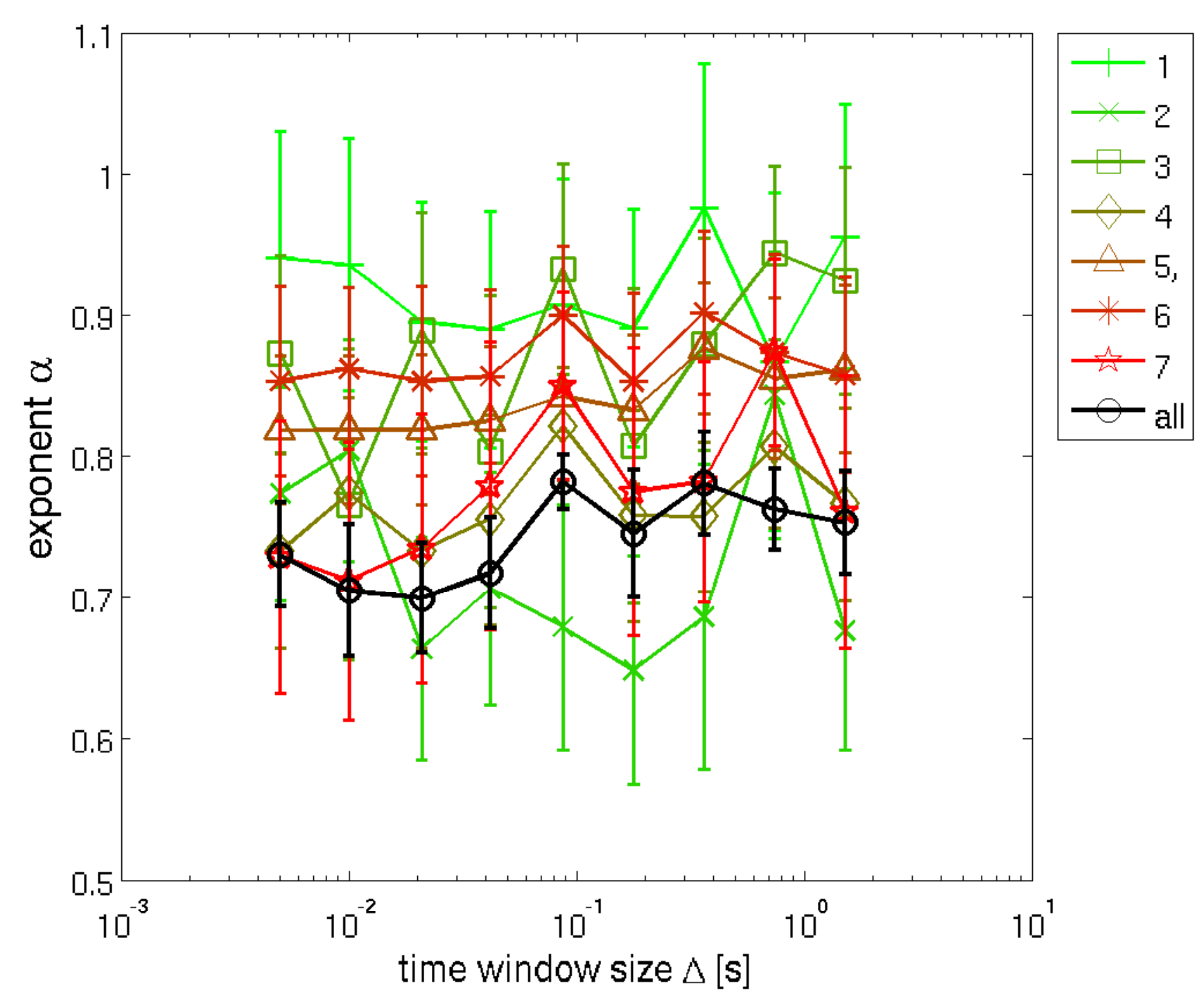}
\caption{\it PHSC (valence)}
\label{fig:alfa-phsc-val}
\end{subfigure}
\begin{subfigure}[b]{0.45\textwidth}
\centering
\includegraphics[width=\textwidth]{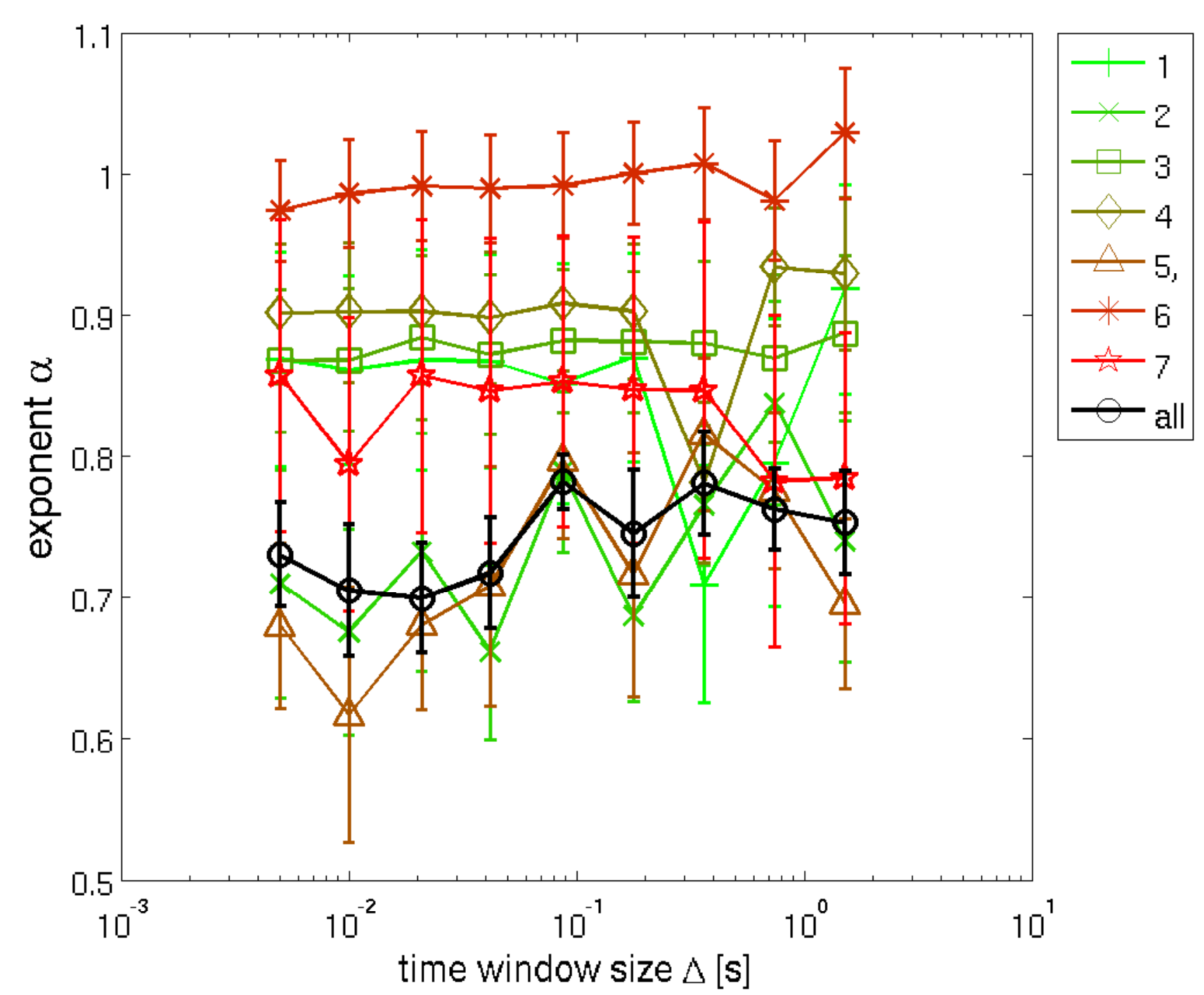}
\caption{\it PHSC (arousal)}
\label{fig:alfa-phsc-aro}
\end{subfigure}
\caption{Temporal fluctuation scaling exponent $\alpha(\Delta)$ for (a,b)~\textbf{ZYG}, (c,d)~\textbf{COR}, (e,f)~\textbf{PHSC} during IAPS stimulation depending on (left) valence score, (right) \textbf{arousal} score (from 1 to 7) as function of window size $\Delta$; black points -- all IAPS data aggregated.}
\label{fig:alfa}
\end{figure}

In the last part of this Section, we show $\alpha$ exponents values for various sizes of observation window~$\Delta$ (Fig.~\ref{fig:alfa}). In all plots X-axes are logarithms of time windows size $\Delta$ and Y-axes correspond to values of the exponent $\alpha(\Delta)$. Colors of the lines mark questionnaire scores in the same way as in the previous graphs. In the case of ZYG and COR signals (Figs.~\ref{fig:alfa-zyg-val}-~\ref{fig:alfa-cor-aro}), differences between exponents grow as well as do error bars for both valence and arousal scores. PHSC results (Figs.~\ref{fig:alfa-phsc-val} and ~\ref{fig:alfa-phsc-aro}) are comparatively noisy for all considered window sizes. The increase of $\alpha$ standard errors with the growth of $\Delta$ is probably an effect of smaller number of full windows in each IAPS-participant pair, which results in a smaller number of values taken to calculate means in each bin. The number of bins has been kept constant for every $\Delta$ value.

\begin{table}[b]
\centering
\begin{tabular}{cc}
\begin{tabular}{rcccccccr}
  \hline
 $v_{ij}$& 1 & 2 & 3 & 4 & 5 & 6 & 7 & $R^2$\\ 
  \hline
1 &  & $^{*}$ & . & $^{*}$ & . &   &   & .95\\ 
  2 & $^{*}$ &  &   &   &   & $^{*}$ & $^{**}$ & .96\\ 
  3 & . &   &  &   &   & . & $^{*}$ & .98\\ 
  4 & $^{*}$ &   &   &  &   & $^{*}$ & $^{**}$ & .98\\ 
  5 & . &   &   &   &  & $^{*}$ & $^{*}$ & .97\\ 
  6 &   & $^{*}$ & . & $^{*}$ & $^{*}$ &  &   & .98\\ 
  7 &   & $^{**}$ & $^{*}$ & $^{**}$ & $^{*}$ &   &  & .98\\ 
   \hline
\end{tabular}
&
\begin{tabular}{rcccccccr}
  \hline
 $a_{ij}$ & 1 & 2 & 3 & 4 & 5 & 6 & 7 & $R^2$\\ 
  \hline
1 &  &   &   &   &   &   & . & .97\\ 
  2 &   &  &   &   &   &   &   & .97\\ 
  3 &   &   &  &   &   &   &   & .98\\ 
  4 &   &   &   &  &   &   & $^{*}$ & .98\\ 
  5 &   &   &   &   &  &   &   & .97\\ 
  6 &   &   &   &   &   &  &   & .97\\ 
  7 & . &   &   & $^{*}$ &   &   &  & .92\\ 
   \hline
\end{tabular}\\
\end{tabular}
\caption{Comparison among pairs of $\alpha$ exponents for different values of valence (left) and arousal (right) in ZYG signal ($\Delta = 0.04$s). Significance codes are the same as in Table \ref{stat:ancova}. The table is deliberately symmetrical to enhance comparison feasibility. The last column contains coefficient of determination $R^2$ values for corresponding fits from Figs.~\ref{fig:tfs-iaps-zyg-val} and ~\ref{fig:tfs-iaps-zyg-aro}.}  
\label{stat:zyg}
\end{table}

\begin{table}[b]
\centering
\begin{tabular}{cc}
\begin{tabular}{rcccccccr}
  \hline
 $v_{ij}$ & 1 & 2 & 3 & 4 & 5 & 6 & 7 & $R^2$ \\ 
  \hline
1 &  &   &   &   &   &   & $^{*}$ & .85\\ 
  2 &   &  &   &   &   &   &   & .82\\ 
  3 &   &   &  &   &   &   & $^{*}$ & .86\\ 
  4 &   &   &   &  &   &   & $^{*}$ & .90\\ 
  5 &   &   &   &   &  &   & $^{*}$ & .87\\ 
  6 &   &   &   &   &   &  & $^{*}$ & .84\\ 
  7 & $^{*}$ &   & $^{*}$ & $^{*}$ & $^{*}$ & $^{*}$ &  & .61\\ 
   \hline

\end{tabular}
&
\begin{tabular}{rcccccccr}
  \hline
 $a_{ij}$ & 1 & 2 & 3 & 4 & 5 & 6 & 7 & $R^2$ \\ 
  \hline
1 &  &   &   &   &   & $^{*}$ &   & .76\\ 
  2 &   &  &   &   &   &   &   & .87\\ 
  3 &   &   &  &   &   & . &   & .91\\ 
  4 &   &   &   &  &   &   &   & .84\\ 
  5 &   &   &   &   &  &   &   & .87\\ 
  6 & $^{*}$ &   & . &   &   &  &   & .77\\ 
  7 &   &   &   &   &   &   &  & .75\\ 
  \hline
\end{tabular}\\
\end{tabular}
\caption{Comparison among pairs of $\alpha$ exponents for different values of valence (left) and arousal (right) in COR signal ($\Delta = 0.36$s). Significance codes are the same as in Table \ref{stat:ancova}. The table is deliberately symmetrical to enhance comparison feasibility. The last column contains coefficient of determination $R^2$ values for corresponding fits from Figs.~\ref{fig:tfs-iaps-cor-val} and ~\ref{fig:tfs-iaps-cor-aro}.}  
\label{stat:cor}
\end{table}


\begin{table}[b]
\centering
\begin{tabular}{rcccccccr}
  \hline
 $a_{ij}$ & 1 & 2 & 3 & 4 & 5 & 6 & 7 & $R^2$\\ 
  \hline
1 &  &   &   &   &   &   &   & .88 \\ 
  2 &   &  & . & . &   & $^{*}$ &   & .85 \\ 
  3 &   & . &  &   &   &   &   & .92 \\ 
  4 &   & . &   &  &   &   &   & .94 \\ 
  5 &   &   &   &   &  & $^{*}$ &   & .76 \\ 
  6 &   & $^{*}$ &   &   & $^{*}$ &  &   & .97 \\ 
  7 &   &   &   &   &   &   &  & .78 \\ 
   \hline
\end{tabular}
\caption{Comparison among pairs of $\alpha$ exponents for different values of arousal in PHSC signal ($\Delta = 0.04$s). Significance codes are the same as in Table \ref{stat:ancova}. The table is deliberately symmetrical to enhance comparison feasibility. The last column contains coefficient of determination $R^2$ values for corresponding fits from Fig.~\ref{fig:tfs-iaps-phsc}.}  
\label{stat:phsc}
\end{table}

\FloatBarrier

\section{Conclusions}

In this paper, we investigate scaling of fluctuations in signals of psychophysiological human activity (facial muscles, skin conductance) that were elicited in response to emotional stimuli. The experiment conducted at Jacobs University with over 60 participants allowed us to test the hypothesis that emotional states can be recognized by examining the scaling exponent of the relation between the standard deviation and the mean.

We underline that our primary observation in this study is that of the existence of Taylor's scaling in the whole time series, where we observed scaling exponents $\alpha$ ranging from $0.83$ to $1.03$. Such values, that are significantly higher than $\alpha = 0.5$ lead us to speculate on the origin of the influence exerted on participants. As it  is well known, the value of $\alpha=0.5$ for temporal fluctuations scaling  is found  for  example \cite{kertesz} in  systems consisting of non-interacting elements. Larger values of  $\alpha$ are observed when interactions take place and elements are  partially synchronized, and/or there is an external impact acting on the system. The results obtained in this paper can not exclude any of these scenarios, since muscle fibers {\it are} interacting with one another, and there is also a complex influence of external emotional signals on these muscles. While the standard assumption in the field of psychophysiological research on facial EMG \cite{tassinary, boxtel,hof} has been that of a quasi-random firing of muscle units (MUs), our results are consistent with more recent findings \cite{deluca}, suggesting a certain degree of synchronization of MU firings.

The second part of our analysis was devoted to separating scaling relations for different levels of subjectively reported emotional valence and arousal that were elicited in participants by exposing them to (emotionally) standardized pictures, as reflected by the questionnaire data. Based on values of scaling exponents obtained by grouping the series with similar emotional scoring, we are able to distinguish series connected to extreme emotions. Interestingly, the results for facial activity ({\it zygomaticus major}, responsible for smiling, and {\it corrugator supercilii} that controls frowning) show that time series for very positive and highly aroused levels are described by low (in comparison to those for different levels) scaling exponents. We speculate that this kind of emotional impact leads to a decrease of the coherent character of the motion of face muscles, which would lower the value of $\alpha$ exponent.   

In the last part of our study, we examine closely the issue of the size of the time window and its influence on the scaling exponent. We observe high variability and almost a monotonic growth of $\alpha$ with increasing length of the time window for the smiling muscle, regardless of the valence and arousal level. Such a behavior can be consistent with the concept of internal synchronization of muscle fibers that should be more easily observed at longer time scales. A similar influence of time window length was reported for volumes of transactions for stocks at NASDAQ, NYSE and  Chinese stock  markets \cite{kertesz,eisler}. In the case of {\it corrugator supercilii}, we deal with an opposite situation of exponent value decay with growing $\Delta$ that might be related to fundamental differences in the interplay between short-term bursts of activity at this site (brief episodes of frowning), and more long-term shifts in the overall tension found at this site for some subjects. 

Our results might contribute to the development of novel approaches to fEMG and EDA signal analysis but still require additional analyses and replication. In future research, we plan to use various detrending algorithms to remove possible effects of data non-stationarity as well to combine the Taylor's studies with the Hurst exponent analysis \cite{kertesz}. 

\section{Acknowledgments}
The research leading to these results has received funding from the EU Seventh Framework Programme (FP7/2007-2013) under grant agreement no. 231323 (\textit{Collective Emotions in Cyberspace} project -- CyberEMOTIONS). J.Ch, A.Ch. J.S. and J.A.H. acknowledge support from Polish Ministry of Science Grant 1029/7.PR UE/2009/7.
J.A.H. has also been partially supported by the Russian Scientific Foundation, proposal \#14-21-00137. We thank Elena Tsankova, Mathias Theunis, and Aleksandra \'{S}widerska for their support with the data collection and their comments on earlier versions of the data analysis. 
\
\section*{References}

\bibliography{temporal-taylors-scaling}

\end{document}